\documentclass[final,3p]{elsarticle}

\usepackage{graphicx}
\usepackage{dcolumn}
\usepackage{color}
\usepackage{xcolor}
\usepackage[utf8]{inputenc}
\usepackage[T1]{fontenc}

\usepackage{graphicx,epstopdf}
\biboptions{sort&compress}

\usepackage{setspace}
\usepackage{multirow}
\usepackage{amsmath}
\usepackage{amssymb}
\usepackage{amsfonts}
\usepackage{graphicx,epstopdf}
\usepackage{float}

\usepackage{caption, subcaption} 
\usepackage{multirow}
\usepackage{bm}

\begin{document}

\journal{arXiv}
\title{Fractional lower-order covariance-based measures for cyclostationary time series with heavy-tailed distributions: application to dependence testing and model order identification}


\author[label1]{Wojciech Żuławiński}

\author[label1]{Agnieszka Wy{\l}oma{\'n}ska\corref{cor1}}
\ead{agnieszka.wylomanska@pwr.edu.pl}
\cortext[cor1]{Corresponding author.}

\address[label1]{Faculty of Pure and Applied Mathematics, Hugo Steinhaus Center,
Wroclaw University of Science and Technology, 50-370 Wroclaw, Poland}

\date{\today}

\begin{abstract}
This article introduces new methods for the analysis of cyclostationary time series with infinite variance. Traditional cyclostationary analysis, based on periodically correlated (PC) processes, relies on the autocovariance function (ACVF). However, the ACVF is not suitable for data exhibiting a heavy-tailed distribution, particularly with infinite variance. Thus, we propose a novel framework for the analysis of cyclostationary time series with heavy-tailed distribution, utilizing the fractional lower-order covariance (FLOC) as an alternative to covariance. This leads to the introduction of two new autodependence measures: the periodic fractional lower-order autocorrelation function (peFLOACF) and the periodic fractional lower-order partial autocorrelation function (peFLOPACF). These measures generalize the classical periodic autocorrelation function (peACF) and periodic partial autocorrelation function (pePACF), offering robust tools for analyzing infinite-variance processes. Two practical applications of the proposed measures are explored: a portmanteau test for testing dependence in cyclostationary series and a method for order identification in periodic autoregressive (PAR) and periodic moving average (PMA) models with infinite variance. Both applications demonstrate the potential of  new tools, with simulations validating their efficiency. The methodology is further illustrated through the analysis of real-world air pollution data, which showcases its practical utility. The results indicate that the proposed measures based on FLOC  provide reliable and efficient techniques for analyzing cyclostationary processes with heavy-tailed distributions.
\end{abstract}

\maketitle

\section{Introduction}

Cyclostationary time series, established in the seminal works of Gudzenko \cite{gudzenko1959pc} and Gladyshev \cite{gladyshev1961pc}, constitute one of the most important classes of non-stationary stochastic processes. In general, the cyclostationarity property refers to periodic variations of some statistical characteristics of a given process. Since many real-world phenomena are inherently random and periodic, cyclostationary processes have gained great interest in various practical applications, e.g., in condition monitoring \cite{antoni2007cyclic,antoni2007cyclic2,feng2021cyclo,yu2013new}, communications \cite{gardner1994communications,nouri2021_5g}, finance \cite{broszkiewicz2004detecting}, hydrology \cite{vecchia1985periodic, mondal2006generating,treistman2020periodic}, malware detection \cite{nkongolo2022application} and environmental engineering \cite{solci2020empirical,sarnaglia2021mregression}.

The classical theory of cyclostationarity is devoted to periodically correlated (PC) processes, i.e., those for which the mean and autocovariance (ACVF) functions are periodic. In this case, we are dealing with the so-called second-order cyclostationary processes.  Although the theory and methods of PC processes have been well developed \cite{hurd2007periodically,napolitano2012book,napolitano2019cyclostationary}, this class is not always suitable for practical modeling of periodically structured phenomena. In particular, when the analyzed data contain outliers, classical approaches based on PC processes may fail. In such cases, one can instead use models based on heavy-tailed distributions, that is, those for which the probability of values of large magnitude is higher than in the Gaussian distribution. In this class, one can distinguish distributions with infinite variance, such as the $\alpha$-stable distribution \cite{samorodnitsky1994stable,weron1994stable,nolan2020stable} which generalizes the Gaussian distribution and is significant from a practical point of view due to the generalized central limit theorem. The $\alpha$-stable distribution has been used for modeling in various fields, e.g., in condition monitoring \cite{yu2013new,zak2019periodically}, finance \cite{mcculloch1996thirteen}, physics \cite{takayasu1984stable}, signal processing \cite{nikias1995signal}, (including audio \cite{leglaive2019speech}, image \cite{gao2023image} and graph \cite{nguyen2020graph} signal processing) and deep learning \cite{yuan2025neural}.

An obvious drawback of infinite-variance distributions is that methods based on covariance/ACVF cannot be used in such a case. In particular, an infinite-variance process with periodic structure cannot be described using the PC property. Instead, cyclostationarity must be redefined to refer to periodic variations of other statistical characteristics which are valid for infinite-variance time series, e.g., covariation \cite{kruczek2021generalized} or codifference \cite{kruczek2020detect} which are popular alternatives to covariance. In this article, we use another measure suitable for the infinite-variance case, namely the fractional lower-order covariance (FLOC) \cite{ma1996joint}, considering a variant of cyclostationarity (referred to as FLOC-cyclostationarity) in which the mean and the FLOC-based counterpart of the ACVF vary periodically. 

In the literature, many examples can be found in which FLOC was successfully applied as an efficient alternative to covariance \cite{bian2010polinsar,feng2023active,zak2019periodically,chen2016floc}. In the cyclostationary analysis, it has been considered mainly in the frequency domain \cite{liu2012tdoa,liu2015joint,liu2018improved,zhao2010doa,ma2010flocs}. In our previous works, FLOC was used to design time domain-based methods (e.g., for estimation) devoted to particular cyclostationary models \cite{zulawinski2021alternative,zulawinski2023eusipco,zulawinski2024eusipco,giri2021floc}. In this article, we continue this research by extending the time-domain methodology for the analysis of FLOC-cyclostationary time series. We propose two FLOC-based autodependence measures -- periodic fractional lower-order autocorrelation function (peFLOACF) and periodic fractional lower-order partial autocorrelation function (peFLOPACF). They are generalizations of the covariance-based measures well established for PC processes -- periodic autocorrelation function (peACF) and periodic partial autocorrelation function (pePACF). The definitions of the proposed measures are very similar to the classical ones, following the fact that FLOC generalizes covariance. Hence, the peFLOACF and peFLOPACF have similar properties for FLOC-cyclostationary models as the peACF and pePACF for PC time series. 

Based on this relation, one can design new methods of the FLOC-cyclostationary time series analysis by replacing the classical measures with the novel ones in procedures developed for PC processes. In this article, we present two such examples of applications of the peFLOACF and peFLOPACF. First, we present a portmanteau test based on the peFLOACF for testing of dependence in FLOC-cyclostationary time series. It is an adaptation of the peACF-based test presented in \cite{mcleod1994diagnostic} (which itself is a generalization of the classical Box-Pierce \cite{box1970test} and Ljung-Box \cite{ljung1978test} tests). The second example is related to the order identification in periodic autoregressive (PAR) and periodic moving average (PMA) models in infinite-variance case. Both time series are special cases of the periodic autoregressive moving average (PARMA) model, a generalization (with periodic parameters) of the well-known autoregressive moving average (ARMA) time series. In its finite-variance version \cite{jones1967par,vecchia1985periodic}, the PARMA model is one of the basic examples of a PC process. In this case, pePACF can be used to identify the order of PAR model \cite{hipel1994book} and peACF -- to estimate the order of PMA time series \cite{ula2003pma}, because of the unique relation between these measures and the corresponding models ("cut-off" property). These techniques are based on the methods well known in classical time series analysis which utilize partial autocorrelation and autocorrelation, respectively, in the order identification for autoregressive (AR) and moving average (MA) models \cite{box2015tsa}. Analogously, as presented in this article, peFLOPACF can be used to determine the order of PAR model, and peFLOACF to identify the PMA model order for FLOC-cyclostationary versions of these time series. Since infinite-variance PARMA models have been widely considered in the literature \cite{nowicka2006dependence,zulawinski2021alternative, kruczek2017modified}, these new order identification procedures are of great practical interest. 

Both presented applications of the introduced measures are also based on previous research related to applications of alternative dependence measures in the analysis of stationary processes (where stationarity was expressed using these alternatives instead of ACVF). In \cite{gallagher2006portmanteau} and \cite{rosadi2009portmanteau}, respectively, portmanteau tests based on covariation and codifference are presented. Covariation was also used for order identification in AR \cite{balakrishna2012stable} and MA \cite{gallagher2002order} models.

As indicated by the performed simulation study, the new measures and methods based on them yield an acceptable efficiency and can be reliable tools for the analysis of cyclostationary time series with heavy tails. This is also shown in the presented analysis of air pollution data using the presented methodology.

The paper is structured as follows. In Section 2, the analyzed class of stochastic processes -- FLOC-cyclostationary time series -- is described. For the analysis of such models, in Section 3, two new autodependence measures are introduced -- the peFLOACF and peFLOPACF. Section 4 presents two examples of the application of the proposed functions, in the portmanteau test for testing of dependence and in the order identification procedure for PAR/PMA models. In Section 5, both methods are validated using Monte Carlo simulations. In Section 6, we present the application of the introduced methodology in the real data analysis. Section 7 concludes the paper.

\subsection*{The following nomenclature is used:}

PC -- periodically correlated

FLOC -- fractional lower-order covariance

FLOM -- fractional lower-order moment

ACVF -- autocovariance function

FLOACVF -- fractional lower-order autocovariance function

peWN -- periodic white noise

peFLOWN -- periodic fractional lower-order white noise

i.i.d. -- independent identically distributed

i.p.d. -- independent periodically distributed

PARMA -- periodic autoregressive moving average

PAR -- periodic autoregressive

PMA -- periodic moving average

peACVF -- periodic autocovariance function $\gamma_v(h)$

peFLOACVF -- periodic fractional lower-order autocovariance function $\psi_v(h)$

peACF -- periodic autocorrelation function $\rho_v(h)$

peFLOACF -- periodic fractional lower-order autocorrelation function $\eta_v(h)$

pePACF -- periodic partial autocorrelation function $\beta_v(h)$

peFLOPACF -- periodic fractional lower-order partial autocorrelation function $\zeta_v(h)$

\section{Cyclostationary models with heavy-tailed distribution}

A stochastic process is called cyclostationary if some of its statistical characteristics vary in a periodic manner. Usually, this term is used to describe the class of PC processes which are characterized by periodicity of the mean and ACVF. However, since this definition is not valid for infinite-variance processes, we consider another type of cyclostationarity that is based on the FLOC. Such processes are hereinafter called FLOC-cyclostationary. In this section, we introduce this concept, as well as the selected models belonging to this class. In this article, we only consider discrete-time stochastic processes (time series).

\subsection{FLOC-cyclostationary time series}

A finite-variance time series $\{X_t\}$, $t\in\mathbb{Z}$, is PC if its mean and ACVF are periodic in $t$ with the same period $T$, that is, if for all $t,h\in\mathbb{Z}$ we have
\begin{equation}
    \mathbb{E}X_t=\mathbb{E} X_{t+T},\qquad
\text{Cov}(X_t,X_{t+h})=\text{Cov}(X_{t+T},X_{t+h+T}).
\end{equation}
PC time series are also called second-order cyclostationary. However, as mentioned, this property is not well defined for infinite-variance time series. To address this issue, the cyclostationarity for this class can be defined similarly as in the second-order case, but with covariance replaced by another dependence measure (that is well defined for infinite-variance time series). In this article, we use the FLOC for this purpose. This dependence measure is a generalization of the classical covariance and was originally proposed for $\alpha$-stable distributions \cite{ma1996joint}. However, because of its simple moment-based form, it can also be applied to other classes (in particular for infinite-variance cases). For two random variables $Y_1,Y_2$ that have finite fractional moments up to some order $1<a\leq 2$ ($\mathbb{E}|Y_1|^r<\infty$, $\mathbb{E}|Y_2|^r<\infty$ for $0<r<a$), FLOC is defined as 
\begin{equation}\label{eq:floc}
    \text{FLOC}(Y_1,Y_2;A,B) = \mathbb{E}[Y_1^{<A>} Y_2^{<B>}],
\end{equation}
 where ${}^{<\cdot>}$ is the signed power ($x^{<c>} = |x|^c \text{ sgn } x$), and the powers $A,B>0$ must satisfy the condition $A+B<a$. Although we restrict ourselves to $1<a\leq 2$ (for further considerations), let us mention that FLOC can also be defined for a more general assumption of $0<a\leq 2$ (e.g. even for random variables with infinite mean). Some properties of the FLOC measure can be found, e.g., in \cite{kruczek2019fractional}. Here, we recall those that are most important in the context of the presented work. For arbitrary $a,b\in\mathbb{R}$, we have
\begin{equation}\label{eq:floc_quasilinear}
    \text{FLOC}(aY_1,bY_2;A,B) = a^{<A>}b^{<B>}\text{FLOC}(Y_1,Y_2;A,B).
\end{equation}
Moreover, if $Y_1$ and $Y_2$ are independent, then 
\begin{equation}\label{eq:floc_independent}
    \text{FLOC}(Y_1,Y_2;A,B) = 0.
\end{equation}
However, unlike covariance, in general, FLOC is not a symmetric measure; that is, $\text{FLOC}(Y_1,Y_2;A,B) \neq \text{FLOC}(Y_2,Y_1;A,B)$ (unless $A=B$). 

We can also consider the result of taking FLOC of a random variable with itself. This quantity, further referred to as fractional lower-order moment (FLOM), for a random variable $Y_1$ has the following form
\begin{equation}
    \text{FLOM}(Y_1;A,B) = \text{FLOC}(Y_1,Y_1;A,B) = \mathbb{E}|Y_1|^{A+B}.
\end{equation}
In the literature, FLOMs are often used to measure the dispersion of infinite-variance random variables, especially for $\alpha$-stable distributions \cite{nikias1995signal}. Here, we consider FLOM as a generalization of variance that will be useful in further considerations; FLOM is the FLOC of a random variable with itself just as the variance is the covariance of a random variable with itself.

In the same way as the covariance is used to define the ACVF, we can use the FLOC measure to construct the fractional lower-order autocovariance function (FLOACVF) which for a time series $\{X_t\}$ is defined as
\begin{equation}\label{eq:autofloc}
    \text{FLOC}(X_t,X_{t+h};A,B) = \mathbb{E}[X_t^{<A>} X_{t+h}^{<B>}],
\end{equation}
when $\{X_t\}$ has finite fractional moments up to some order $1<a\leq 2$ (for each $t\in\mathbb{Z}$, $\mathbb{E}|X_t|^r<\infty$ for $0<r<a$), and the condition $A+B<a$ is satisfied. When $A=B=1$, the FLOACVF is equal to the ACVF (consequently, FLOC is then equal to covariance, and FLOM -- to variance).

Finally, using the FLOACVF, we can define the property that is analogous to the PC but is also valid for infinite-variance time series. A time series $\{X_t\}$ (for which given FLOC exists) is said to be FLOC-cyclostationary, if its mean and FLOACVF are periodic in $t$ with the same period $T$, i.e., for all $t,h \in \mathbb{Z}$ we have
\begin{equation}
    \mathbb{E}X_t=\mathbb{E}X_{t+T},\qquad
\text{FLOC}(X_t,X_{t+h};A,B)=\text{FLOC}(X_{t+T},X_{t+h+T};A,B),
\end{equation}
for given $A,B$ applied in the FLOACVF.

In the following sections, we introduce the FLOC-cyclostationary time series, which will be considered in further parts. Henceforth, we assume that all the {time series} considered have mean equal to 0.

\subsection{Periodic fractional lower-order white noise}\label{subsec:peflown}

One of the simplest examples of finite-variance weakly stationary (later called stationary) time series is the white noise, which is a sequence of uncorrelated zero-mean random variables with constant variance. Analogously, as an example of PC time series, we can consider the periodic white noise (peWN) \cite{boshnakov1995parma}. A finite-variance time series $\{X_t\}$ is peWN with period $T$ if, for each $t\in\mathbb{Z}$ and $h\in\mathbb{Z} \setminus \{0\} $, we have
\begin{equation}\label{eq:pewn}
    \mathbb{E}X_t = 0,\quad \text{Var}(X_t) = \text{Var}(X_{t+T}), \quad \text{Cov}(X_t,X_{t+h}) = 0.
\end{equation}
Let us now generalize the concept of peWN by replacing the variance and covariance in conditions \eqref{eq:pewn} with FLOM and FLOC, respectively. This time series will be called periodic fractional lower-order white noise (peFLOWN). A time series $\{X_t\}$ (possibly with infinite variance, but with considered FLOM/FLOC well defined) is said to be peFLOWN with period $T$ if, for each $t\in\mathbb{Z}$ and $h\in\mathbb{Z} \setminus \{0\} $, we have
\begin{equation}\label{eq:peflown}
    \mathbb{E}X_t = 0,\quad \text{FLOM}(X_t;A,B) = \text{FLOM}(X_{t+T};A,B), \quad \text{FLOC}(X_t,X_{t+h};A,B) = 0.
\end{equation}
From the conditions \eqref{eq:peflown}, one can see that a peFLOWN time series is also FLOC-cyclostationary. The simplest example of peFLOWN time series is a sequence of independent periodically distributed (i.p.d.) random variables. A time series $\{X_t\}$ is said to be an i.p.d. sequence with period $T$ if for all $t\in\mathbb{Z}$ we have
\begin{equation}
    X_{t} \overset{d}{=} X_{t+T},
\end{equation}
where $\overset{d}{=}$ denotes the equality in distribution, and if for each $t\in\mathbb{Z}$ and $h \in \mathbb{Z} \setminus \{0\}$, $X_{t}$ and $X_{t+h}$  are independent. Note that this time series is a generalization of an independent identically distributed (i.i.d.) sequence.

\subsection{PARMA, PAR, PMA models}\label{subsec:parma}

The second considered cyclostationary time series is the PARMA model. It is a generalization of the ARMA model, with parameters let to be periodic functions instead of constants (these models are equivalent for period $T=1$). In the following, we present the definition that allows the considered time series to have infinite variance. A time series $\{X_t\}$ is PARMA$_T$($p,q$) if it is {FLOC-cyclostationary} and if for every $t\in\mathbb{Z}$ we have
    \begin{equation}\label{eq:parma}
X_t - \phi_1(t) X_{t-1} - \ldots - \phi_p(t) X_{t-p} = \xi_t + \theta_1(t) \xi_{t-1} + \ldots + \theta_q(t) \xi_{t-q},
    \end{equation}
    where $\{\xi_t\}$ is {peFLOWN}, and the coefficients $\{\phi_i(t),\, i = 1,\ldots,p\}$, $\{\theta_j(t),\, j = 1,\ldots,q\}$ are periodic in $t$ with the same period $T$. In this article, we consider PARMA models that satisfy the causality condition, that is, for each $t\in\mathbb{Z}$, the following representation exists
    \begin{equation}\label{eq:causal}
        X_{t} = \sum_{j=0}^\infty w_j(t) \xi_{t-j},
    \end{equation}
    with $\sum_{j=0}^\infty |w_j(t)|<\infty$, and $w_j(t)=w_j(t+T)$ for all $t\in\mathbb{Z}$.

A PARMA$_T$($p,0$) time series is called PAR model, denoted as PAR$_T$($p$) and given by the following formula
\begin{equation}\label{eq:par}
X_t - \phi_1(t) X_{t-1} - \ldots - \phi_p(t) X_{t-p} = \xi_t.
    \end{equation}
The PMA model is an alternative name for PARMA$_T$($0,q$) case. This time series, denoted as PMA$_T$($q$), is then characterized by
    \begin{equation}\label{eq:pma}
X_t = \xi_t + \theta_1(t) \xi_{t-1} + \ldots + \theta_q(t) \xi_{t-q}.
    \end{equation}
Note that the PAR and PMA time series are analogous to the AR and MA models, which are special cases of the ARMA model. The parameters $p,q$ in the models introduced above are referred to as orders. In practical applications, their identification for the analyzed dataset is one of the crucial steps. In the assumed convention, $p$ and $q$ can be considered as "global" orders of a given model. In addition, let us introduce the notation of local orders $p(t),q(t)$, which can be considered as "actual" orders of a given model at time $t$. For $\{X_t\}$ being a PARMA$_T(p,q)$ model satisfying \eqref{eq:parma}, for given $t\in\mathbb{Z}$, we set
\begin{equation}
    p(t) = \max\{i\: : \: i=1,\ldots,p, \, \phi_i(t) \neq 0\},
\end{equation}
\begin{equation}
    q(t) = \max\{j\: : \: j=1,\ldots,q, \, \theta_j(t) \neq 0\}.
\end{equation}
In other words, $p(t)$ and $q(t)$ are the local orders of PARMA model at time $t$, if, for this $t$,  the formula \eqref{eq:parma} can be equivalently rewritten as 
\begin{equation}
    X_t - \phi_1(t) X_{t-1} - \ldots - \phi_{p(t)}(t) X_{t-p(t)} = \xi_t + \theta_1(t) \xi_{t-1} + \ldots + \theta_{q(t)}(t) \xi_{t-q(t)}.
\end{equation}
Obviously, from the periodicity of the $\phi_i(t)$ and $\theta_j(t)$ coefficients, $p(t)$ and $q(t)$ are also periodic in $t$ with period $T$. Hence, it is sufficient to consider only the sequences $p(1),\ldots,p(T)$ and $q(1),\ldots,q(T)$, further referred to as seasonal orders (local orders for $v=1,\ldots,T$). The "global" orders of a given model can be obtained by setting $p=\max_{t\in\mathbb{Z}} p(t)$, $q=\max_{t\in\mathbb{Z}} q(t)$.

\subsection{Symmetric $\alpha$-stable distribution}
\label{subsec:dists}

In the presented FLOC measure and the introduced cyclostationary time series models, there are no assumptions of specific distributions (apart from the required existence of fractional moments). In particular, as this is the main motivation for applying the FLOC, the time series under consideration are allowed to have infinite variance; hence, the proposed methodology is very general. For illustrative purposes, in this article the considered models are based on the symmetric $\alpha$-stable distribution, a well-known example of a heavy-tailed distribution with infinite variance.

A random variable $Z$ has a symmetric $\alpha$-stable distribution (denoted as $\mathcal{S}(\alpha,\sigma)$), if its characteristic function has the following form
\begin{equation}
   \Phi_Z(s) = \mathbb{E}\exp(isZ) = \exp(-\sigma^\alpha |s|^\alpha),\quad s\in\mathbb{R},
\end{equation}
for the stability index $0<\alpha\leq 2$ and the scale parameter $\sigma>0$. The value of $\alpha$ is crucial for the behavior of this distribution; the lower it is, the heavier the tails are. Moreover, for any $\alpha<2$, only the moments of order smaller than $\alpha$ are finite ($\mathbb{E}|Z|^r < \infty$ for $r<\alpha$, $\mathbb{E}|Z|^r = \infty$ for $r \geq \alpha$). Note that this fact implies that for all $\alpha<2$ the variance of a $\mathcal{S}(\alpha,\sigma)$ distribution is infinite. For $\alpha=2$, this distribution is equivalent to the Gaussian case; hence, usually only the distributions with $\alpha<2$ are considered. If the FLOC between two symmetric $\alpha$-stable random variables with the same $\alpha$ is calculated, then the condition related to the $A,B$ parameters has the form $A+B<\alpha$, which follows from the mentioned property of fractional moments of this distribution.

Let $\{X_t\}$ be a PARMA$_T(p,q)$ with innovations $\{\xi_t\}$ being an i.p.d. sequence from symmetric $\alpha$-stable distribution (with constant $\alpha$), for which the representation \eqref{eq:causal} exists; i.e., for each $t\in\mathbb{Z}$, $X_t$ is an infinite linear combination of symmetric $\alpha$-stable random variables with this $\alpha$ parameter. Then, $X_t$ is also symmetric $\alpha$-stable with the same $\alpha$ \cite{samorodnitsky1994stable}.  Hence, for a FLOC between any pair of PARMA time series values, the condition for the $A,B$ parameters is again $A+B<\alpha$. 

\begin{figure}
    \centering
    \includegraphics[width=0.32\linewidth]{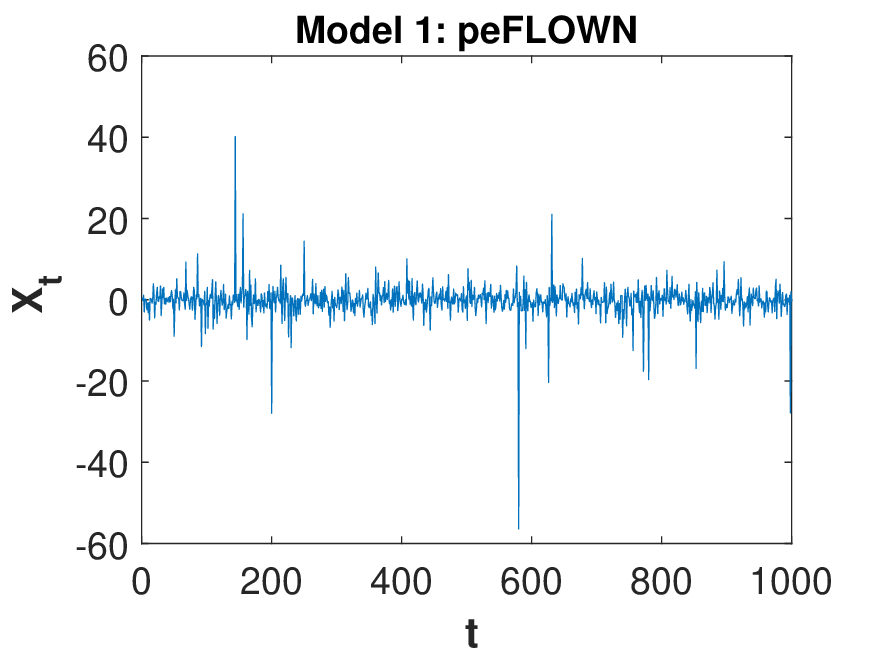}
    \includegraphics[width=0.32\linewidth]{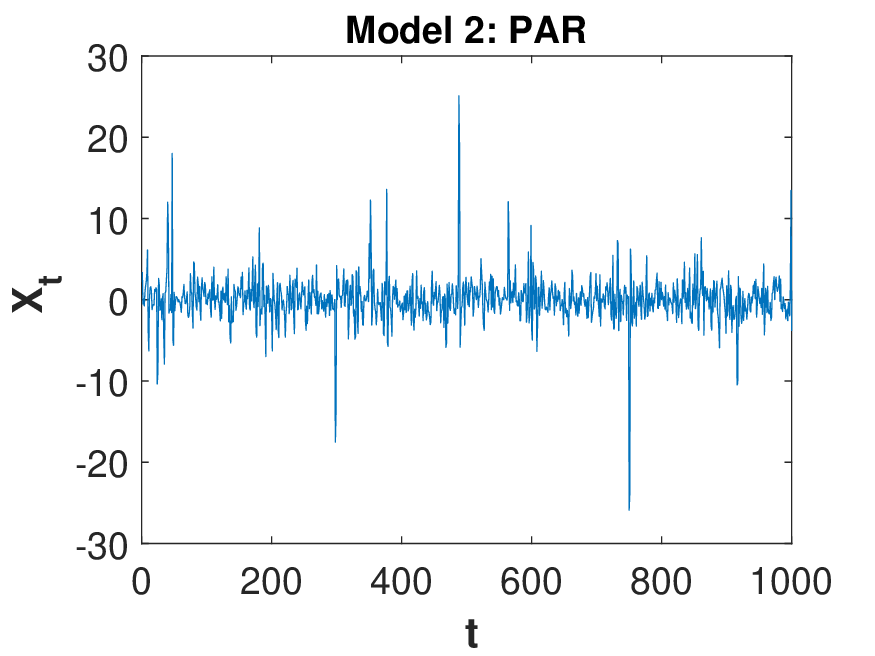}
    \includegraphics[width=0.32\linewidth]{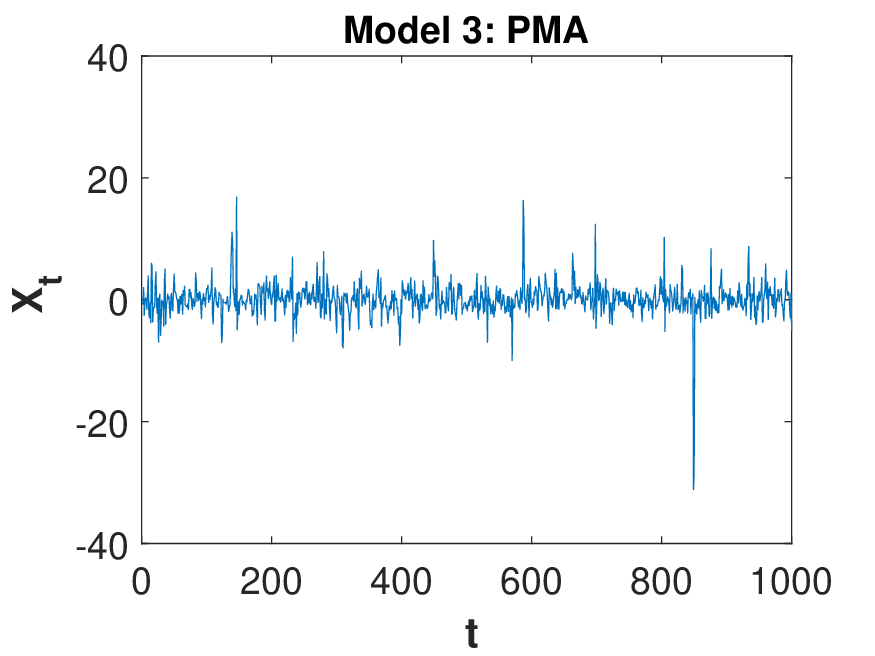}
    \caption{Sample trajectories of selected cyclostationary models with period $T=2$ based on $\mathcal{S}(\alpha=1.7,\sigma)$ distributions: peFLOWN (Model 1), PAR$_2(1)$ (Model 2) and PMA$_2(1)$ (Model 3). Complete specification of each model can be found in Section \ref{subsec:dists}.}
    \label{fig:sample_trajs_stable}
\end{figure}

In Fig. \ref{fig:sample_trajs_stable}, we illustrate sample trajectories of selected cyclostationary models (with period $T=2$ and length $NT=1000$) introduced in Sections \ref{subsec:peflown} and \ref{subsec:parma}. We consider the following three cases of the $\{X_t\}$ time series:
\begin{itemize}
    \item Model 1 (left panel): i.p.d. sequence of $\mathcal{S}(\alpha=1.7,\sigma(t))$ random variables with periodic $\sigma(t)$: $\sigma(1) = 1$, $\sigma(2) = 2$ (example of a peFLOWN sequence). 
    \item Model 2 (middle panel): PAR$_2(1)$ model with $\phi_1(1)=0.8$, $\phi_1(2)=-0.3$, and i.i.d. innovations $\{\xi_t\} \sim \mathcal{S}(\alpha=1.7,\sigma=1)$.
    \item Model 3 (right panel): PMA$_2(1)$ model with $\theta_1(1)=0.8$, $\theta_1(2)=-0.3$, and i.i.d. innovations $\{\xi_t\} \sim \mathcal{S}(\alpha=1.7,\sigma=1)$.
\end{itemize}
In the illustrated sample trajectories, most of all, one can observe the heavy-tailed behavior caused by the underlying $\alpha$-stable distribution (i.e., many significant outliers are present). The presented time series plots do not allow one to distinguish between particular models; however, it will be possible to do so using the dependence measures presented in Section \ref{sec:dependence_measures}, as discussed therein.

\section{The peFLOACF and peFLOPACF}\label{sec:dependence_measures}

\subsection{Autodependence measures for PC time series}\label{subsec:classical_measures}

The FLOC-based measures proposed in this article are generalizations of the covariance-based measures used in the analysis of PC time series. Therefore, in this section, we recall these classical measures. 

For PC time series, the notion of periodic ACVF (peACVF) is very convenient. For a PC time series $\{X_t\}$, it is defined as
\begin{equation}\label{eq:peacvf}
    \gamma_v(h) = \text{Cov}(X_{nT+v}, X_{nT+v-h}) = \mathbb{E}[X_{nT+v} X_{nT+v-h}],
\end{equation}
for $n\in\mathbb{Z}$. Note that under this convention, we have $\text{Var}(X_t) = \gamma_t(0)$. One can also consider the normalized version of the peACVF, that is, the peACF, which measures the covariance between the standardized $X_{nT+v}$ and $X_{nT+v-h}$, that is,
\begin{equation}\label{eq:peacf}
    \rho_v(h) = \text{Cov}\left(\frac{X_{nT+v}}{\sqrt{\gamma_v(0)}}, \frac{X_{nT+v-h}}{\sqrt{\gamma_{v-h}(0)}}\right) = \mathbb{E}\left[\left(\frac{X_{nT+v}}{\sqrt{\gamma_v(0)}}\right) \left(\frac{X_{nT+v-h}}{\sqrt{\gamma_{v-h}(0)}}\right)\right] = \frac{\gamma_v(h)}{\sqrt{\gamma_{v}(0)\gamma_{v-h}(0)}}.
\end{equation}
In this context, standardization refers to its default meaning; that is, to transforming a random variable so that it has unit variance, by dividing by the square root of its variance ($\{X_t\}$ time series is already zero-mean).

The peACVF and peACF serve a similar role in the analysis of PC time series as the classical ACVF and autocorrelation function (ACF) for stationary time series. For the latter class, we can also use the partial autocorrelation function (PACF). Similarly, for PC time series, the pePACF can be considered. Just as the PACF is usually considered because of its unique relation with AR models, the pePACF is well known because of its behavior for (finite-variance) PAR time series. In the literature, there are various approaches for introducing pePACF -- in this article, we consider the one used, e.g., in \cite{hipel1994book} that is based on the Yule-Walker equations (since it can be conveniently modified using FLOC, as shown later). These equations establish the connection between the PAR model parameters and its peACVF -- namely, for PAR$_T(p)$ model $\{X_t\}$, for given $v=1,\ldots,T$, the following relation holds
\begin{equation}\label{eq:yw_par}
   \bm{\phi}_{v} = \left({\mathbf{\Gamma}}_{v}\right)^{-1}  {\bm{\gamma}}_{v},
\end{equation}
where ${\mathbf{\Gamma}}_{v}$ (assumed to be non-singular) is a $p \times p$ matrix defined as
\begin{equation}\label{eq:x_loyw_def1}
(\mathbf{\Gamma}_{v})_{i,j} = \gamma_{v-j}(i-j), \quad i,j=1,\ldots,p,
\end{equation}
and
\begin{equation}\label{eq:x_loyw_def3}
\bm{\gamma}_{v} = [\gamma_v(1),\cdots,\gamma_v(p)]'.
\end{equation}
The definition of pePACF given in \cite{hipel1994book}  naturally follows from the Yule-Walker equations. For a PC time series $\{X_t\}$, for $v=1,\ldots,T$ and $h\in\mathbb{N}$, the pePACF $\beta_v(h)$ is defined as the last component of the vector $\bm{\phi}_{v,h}$ given by 
\begin{equation}\label{eq:pepacf}
   \bm{\phi}_{v,h} = \left({\mathbf{\Gamma}}_{v,h}\right)^{-1}  {\bm{\gamma}}_{v,h},
\end{equation}
where ${\mathbf{\Gamma}}_{v,h}$ (assumed to be non-singular) is a $h \times h$ matrix defined as
\begin{equation}\label{eq:x_loyw_def1h}
(\mathbf{\Gamma}_{v,h})_{i,j} = \gamma_{v-j}(i-j), \quad i,j=1,\ldots,h,
\end{equation}
and
\begin{equation}\label{eq:x_loyw_def3h}
\bm{\gamma}_{v,h} = [\gamma_v(1),\cdots,\gamma_v(h)]'.
\end{equation}
In the literature \cite{mondal2006generating,treistman2020periodic}, one can also find a convention in which the peACF $\rho_v(h)$ is used instead of the peACVF $\gamma_v(h)$ in \eqref{eq:x_loyw_def1h} and \eqref{eq:x_loyw_def3h}.

\subsection{The peFLOACF}\label{subsec:pefloacf}

In the infinite-variance case, for the class of FLOC-cyclostationary time series, a natural counterpart of peACVF can be defined using FLOACVF. We introduce the periodic FLOACVF (peFLOACVF) as
\begin{equation}\label{eq:periodic_floacvf}
    \psi_v(h) = \text{FLOC}(X_{nT+v}, X_{nT+v-h};A,B) = \mathbb{E}[X_{nT+v}^{<A>} X_{nT+v-h}^{<B>}].
\end{equation}
Note that in this measure (as well as in other FLOC-based functions considered below), we do not explicitly list the $A,B$ powers as arguments (for conciseness). We assume that wherever an expression based on FLOC is used, some appropriate values of $A,B$ are applied there.

The peFLOACVF has been analyzed in the literature for estimation purposes \cite{zulawinski2021alternative}. However, to the best of the authors' knowledge, a normalized version of this measure that would be a FLOC-based counterpart of the peACF (or even the classical ACF) has not yet been considered. Since such a function would be of significant practical importance (e.g. because of its scale invariance that will be useful later), let us now introduce the peFLOACF. 

Just as the peACF $\rho_v(h)$ is defined as the covariance between the standardized $X_{nT+v}$ and $X_{nT+v-h}$, the peFLOACF is defined as the FLOC between the standardized $X_{nT+v}$ and $X_{nT+v-h}$. However, first, the standardization procedure must be redefined as it cannot be based on variance. Henceforth, this term will refer to a transformation of a random variable after which the FLOM will be equal to 1. For given $t\in\mathbb{Z}$, to obtain the constant $s(t)>0$ by which one must divide $X_t$ to get a random variable with unit FLOM, let us solve for it the following equation
\begin{equation}
    \text{FLOM}\left(\frac{X_t}{s(t)}\right) =  \left(\frac{1}{s(t)}\right)^{A+B} \mathbb{E}|X_t|^{A+B} = \left(\frac{1}{s(t)}\right)^{A+B} \psi_t(0) = 1,
\end{equation}
using the fact that $\text{FLOM}(X_t) = \psi_t(0)$. Finally, we obtain
\begin{equation}
    s(t) = (\psi_t(0))^{\frac{1}{A+B}}.
\end{equation}
Then, as mentioned, the peFLOACF $\eta_v(h)$ is defined as the FLOC between the standardized (in the FLOM sense) $X_{nT+v}$ and $X_{nT+v-h}$, that is
\begin{equation}
    \eta_v(h) = \text{FLOC}\left(\frac{X_{nT+v}}{s(v)}, \frac{X_{nT+v-h}}{s(v-h)};A,B\right) = \left(\frac{1}{s(v)}\right)^A \left(\frac{1}{s(v-h)}\right)^B \psi_v(h),
\end{equation}
which finally yields
\begin{equation}\label{eq:floacf}
    \eta_v(h) = \frac{\psi_v(h)}{\psi_v(0)^{\frac{A}{A+B}} \psi_{v-h}(0)^{\frac{B}{A+B}}}.
\end{equation}
If $A=B$ then \eqref{eq:floacf} simplifies to 
\begin{equation}
\eta_v(h) = \frac{\psi_v(h)}{\sqrt{\psi_{v}(0)\psi_{v-h}(0)}},
\end{equation}
which is a very similar expression to the peACF \eqref{eq:peacf} (and for $A=B=1$, they are equivalent).

An important feature of the peFLOACF is its characteristic behavior for PMA time series. If $\{X_t\}$ is a PMA$_T(q)$ model (with the additional assumption that its innovations $\{\xi_t\}$ are i.p.d.), then for given $v=1,\ldots,T$, for all $|h|>q$ we have $\eta_v(h)=0$. This is caused by the fact that 
\begin{equation}
    X_{nT+v} = \xi_{nT+v} + \theta_1(v) \xi_{nT+v-1} + \ldots + \theta_{q}(v) \xi_{nT+v-q}
\end{equation}
and
\begin{equation}
    X_{nT+v-h} = \xi_{nT+v-h} + \theta_1(v-h) \xi_{nT+v-h-1} + \ldots + \theta_{q}(v-h) \xi_{nT+v-h-q} 
\end{equation}
are independent for $|h|>q$ (as then $X_{nT+v}$ is composed of different terms from $\{\xi_t\}$ than $X_{nT+v-h}$), and thus $\text{FLOC}(X_{nT+v},X_{nT+v-h};A,B)=0$. This "cut-off" property is analogous to the ones known for the ACF and MA models and for the peACF and finite-variance PMA time series. Because of this relation, as shown later, the peFLOACF can be used to identify the order of the PMA model. For clarity, this property was derived using the "global" order $q$, but it also holds if it is replaced by the seasonal $q(v)$ for given $v$ in $\eta_v(h)$.  

Just as the peFLOACF $\eta_v(h)$ is based on the peFLOACVF $\psi_v(h)$, the estimator of the former function (called sample peFLOACF) is also constructed using the empirical version of the latter. For a sample $x_1,\ldots,x_{NT}$ corresponding to the $\{X_t\}$ time series, the sample peFLOACVF is defined as 
\begin{equation}
    \hat{\psi}_v(h) = \frac{1}{N}\sum_{n=l_b}^{r_b} {x_{nT+v}^{<A>} x_{nT+v-h}^{<B>}},
\end{equation}
where
    \begin{equation}\label{eq:lr}
 l_b=\max\left(\left\lceil\frac{1-v}{T}\right\rceil,\left\lceil\frac{1-(v-h)}{T}\right\rceil\right), \quad  r_b=\min\left(\left\lfloor\frac{NT-v}{T}\right\rfloor,\left\lfloor\frac{NT-(v-h)}{T}\right\rfloor\right).
\end{equation}
Then, the sample peFLOACF is given by 
\begin{equation}\label{eq:floacf_est}
        \hat{\eta}_v(h) = \frac{\hat{\psi}_v(h)}{\hat{\psi}_v(0)^{\frac{A}{A+B}} \hat{\psi}_{v-h}(0)^{\frac{B}{A+B}}}.
\end{equation}

\begin{figure}
    \centering
    \includegraphics[width=0.32\linewidth]{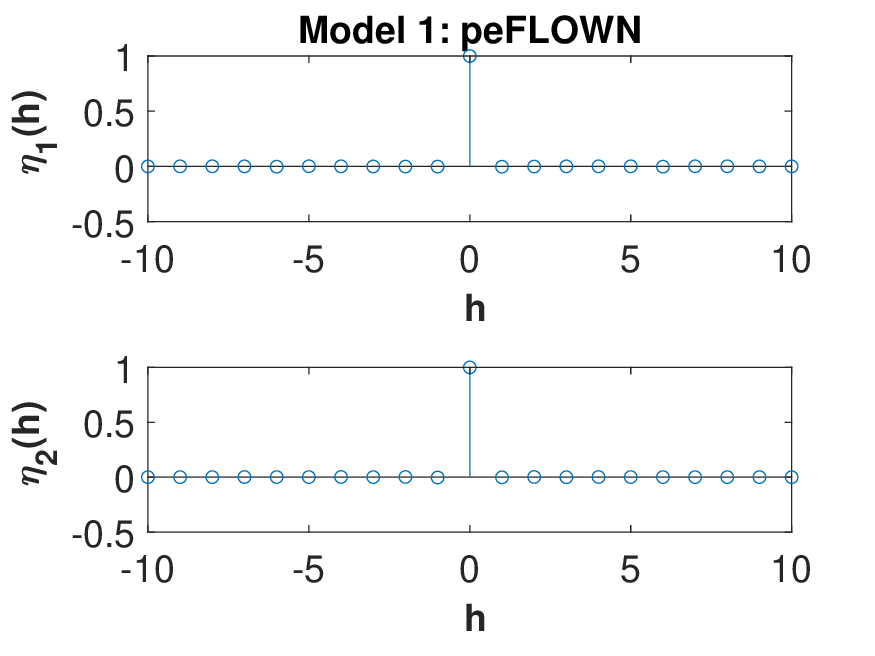}
    \includegraphics[width=0.32\linewidth]{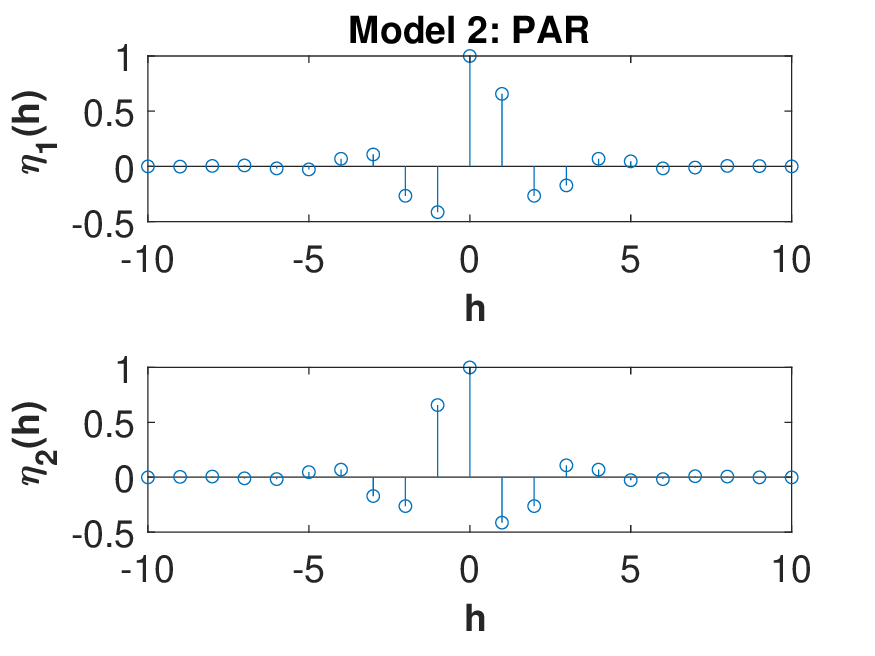}
    \includegraphics[width=0.32\linewidth]{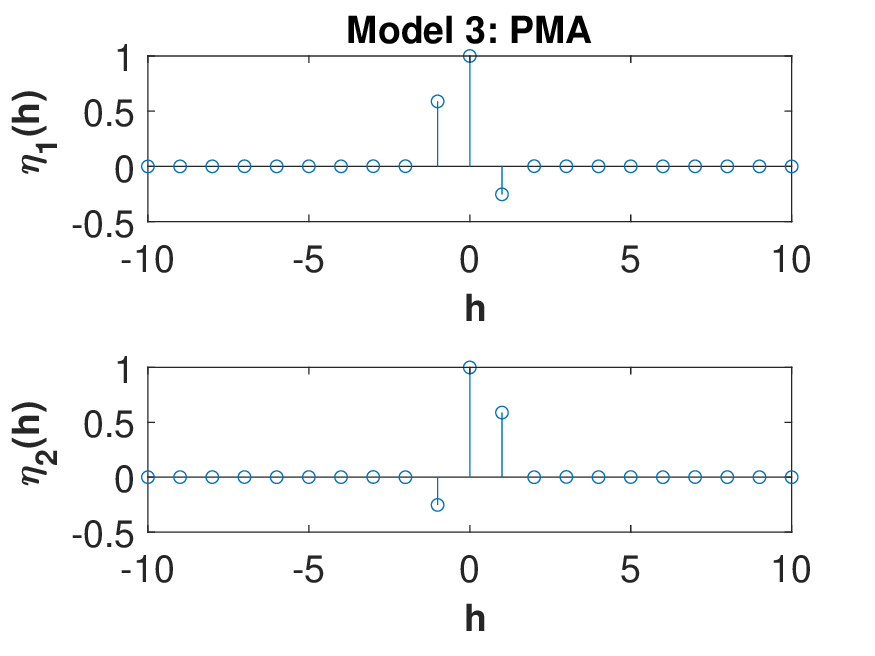}
    \caption{Values of peFLOACF $\eta_v(h)$ with $A=B=0.8$ for $v=1$ (top panels) and $v=2$ (bottom panels) for selected cyclostationary models with period $T=2$: peFLOWN (Model 1), PAR$_2(1)$ (Model 2) and PMA$_2(1)$ (Model 3). Complete specification of each model can be found in Section \ref{subsec:dists}.}
    \label{fig:sample_trajs_pefloacf}
\end{figure}

Let us now illustrate the introduced peFLOACF measure for Models 1-3 considered in Section \ref{subsec:dists} and illustrated in Fig. \ref{fig:sample_trajs_stable}. We assume that $A=B=0.8$, which is valid for the considered time series where $\alpha=1.7$. For each model, the plots of peFLOACF ${\eta}_v(h)$ for $v=1,2$ (recall that $T=2$) and $h=-10,\ldots,10$ are presented in Fig. \ref{fig:sample_trajs_pefloacf}. These results were obtained using sample peFLOACF $\hat{\eta}_v(h)$ averaged over 1000 trajectories of length $NT=1000$ of a given model. The behavior of peFLOACF observed for peFLOWN sequence (Model 1), that is, all values of $\eta_v(h)$ for $h\neq 0$ are equal to 0, follows directly from its definition. For PAR model (Model 2), we observe some non-zero values of peFLOACF, but no particular pattern can be identified, except for decaying to zero for $|h|\rightarrow \infty$. On the other hand, for PMA time series (Model 3), for all $|h|>q$ (recall that here $q=1$), the value of $\eta_v(h)$ is equal to zero. This observation confirms the "cut-off" property of peFLOACF for PMA models (shown above).

\subsection{The peFLOPACF}\label{subsec:peflopacf}

Analogously, the concept of pePACF can be generalized using the FLOC to a similar function suitable for the infinite-variance case, that is, the peFLOPACF.

To construct a FLOC-based counterpart of the pePACF, let us recall that the latter measure was defined using the Yule-Walker equations based on peACVF terms. Analogous Yule-Walker equations based on the peFLOACVF measure, derived in \cite{zulawinski2021alternative}, have the following form -- for PAR$_T(p)$ model $\{X_t\}$, for $v=1,\ldots,T$, the following relation holds
\begin{equation}\label{eq:yw_floc}
   \bm{\phi}_{v} = \left({\mathbf{\Psi}}_{v}\right)^{-1}  {\bm{\psi}}_{v},
\end{equation}
where ${\mathbf{\Psi}}_{v}$ (assumed to be non-singular) is of elements
\begin{equation}\label{eq:x_loyw_def1_floc}
(\mathbf{\Psi}_{v})_{i,j} = \psi_{v-j}(i-j), \quad i,j=1,\ldots,p,
\end{equation}
and
\begin{equation}\label{eq:x_loyw_def3_floc}
\bm{\psi}_{v} = [\psi_v(1),\cdots,\psi_v(h)]'.
\end{equation}
However, from the derivation of \eqref{eq:yw_floc} (which is based on transformations of the PAR equation \eqref{eq:par}, see \cite{zulawinski2021alternative}), we must take $A=1$ in all peFLOACVF terms in \eqref{eq:x_loyw_def1_floc} and \eqref{eq:x_loyw_def3_floc}. We assume this value in all peFLOPACF-related considerations below.

Given the FLOC-based Yule-Walker equations \eqref{eq:yw_floc}, mimicking the presented definition of the pePACF, one can define the peFLOPACF $\zeta_v(h)$ for a PC time series $\{X_t\}$, for $v=1,\ldots, T$, $h\in\mathbb{N}$, as the last component of the vector $\bm{\phi}_{v,h}$ given by 
\begin{equation}\label{eq:pepacf_acvf_floc}
   \bm{\phi}_{v,h} = \left({\mathbf{\Psi}}_{v,h}\right)^{-1}  {\bm{\psi}}_{v,h},
\end{equation}
where ${\mathbf{\Psi}}_{v,h}$ (assumed to be non-singular) is of elements
\begin{equation}\label{eq:x_loyw_def1h_acvf_floc}
(\mathbf{\Psi}_{v,h})_{i,j} = \psi_{v-j}(i-j), \quad i,j=1,\ldots,h,
\end{equation}
and
\begin{equation}\label{eq:x_loyw_def3h_acvf_floc}
\bm{\psi}_{v,h} = [\psi_v(1),\cdots,\psi_v(h)]'.
\end{equation}
Note that for a PAR$_T(p)$ model $\{X_t\}$ and for $h=p$, the equation \eqref{eq:pepacf_acvf_floc} is equivalent to the FLOC-based Yule-Walker equations \eqref{eq:yw_floc}. Hence, in this case, if the peFLOPACF $\zeta_v(h)$ is defined through \eqref{eq:pepacf_acvf_floc}, we have $\zeta_v(p) = \phi_p(v)$. Similar reasoning leads to the conclusion that $\zeta_v(h)=0$ if $h>p$. Note that PAR$_T(p)$ is equal to PAR$_T(p')$ model with $p'>p$, where $\phi_i(v)=0$ for $i=p+1,\ldots,p'$. Then, if we derive the Yule-Walker equations \eqref{eq:yw_floc} using $p'$ as the assumed order, we obtain $\zeta_v(p') = \phi_{p'}(v)=0$. This "cut-off" property is characteristic for "partial" measures; it can also be observed for the PACF in AR models and the pePACF in (finite-variance) PAR time series. Naturally, this feature of the peFLOPACF can be utilized to determine the order of PAR model. For clarity, we used the "global" order $p$ in the properties listed above, but they also hold if $p$ is replaced by the seasonal order $p(v)$ for given $v$ in $\zeta_v(h)$.

As mentioned in Section \ref{subsec:classical_measures}, an alternative definition of the pePACF is based on replacing the peACVF terms in the definition presented in this article with the peACF-based ones. An analogous approach can also be considered in this case, where the peFLOACVF terms in \eqref{eq:x_loyw_def1h_acvf_floc} and \eqref{eq:x_loyw_def3h_acvf_floc} are replaced by the peFLOACF counterparts. Henceforth, we use this method of defining the peFLOPACF (described below), since it proved to be significantly more efficient in the performed experiments.

For a FLOC-cyclostationary time series $\{X_t\}$, we define the peFLOPACF $\zeta_v(h)$, for $v=1,\ldots, T$, $h\in\mathbb{N}$, as the last component of the vector $\bm{\phi}_{v,h}$ given by 
\begin{equation}\label{eq:pepacf_acf_floc}
   \bm{\phi}_{v,h} = \left({\mathbf{H}}_{v,h}\right)^{-1}  {\bm{\eta}}_{v,h},
\end{equation}
where ${\mathbf{H}}_{v,h}$ (assumed to be non-singular) is of elements
\begin{equation}\label{eq:x_loyw_def1h_acf_floc}
(\mathbf{H}_{v,h})_{i,j} = \eta_{v-j}(i-j), \quad i,j=1,\ldots,h,
\end{equation}
and
\begin{equation}\label{eq:x_loyw_def3h_acf_floc}
\bm{\eta}_{v,h} = [\eta_v(1),\cdots,\eta_v(h)]'.
\end{equation}

In practice, the peFLOPACF $\zeta_v(h)$ is estimated through the sample peFLOPACF $\hat{\zeta}_v(h)$, where all $\eta_v(h)$ expressions in \eqref{eq:x_loyw_def1h_acf_floc} and \eqref{eq:x_loyw_def3h_acf_floc} are replaced by their empirical versions $\hat{\eta}_v(h)$ \eqref{eq:floacf_est}.

\begin{figure}
    \centering
    \includegraphics[width=0.32\linewidth]{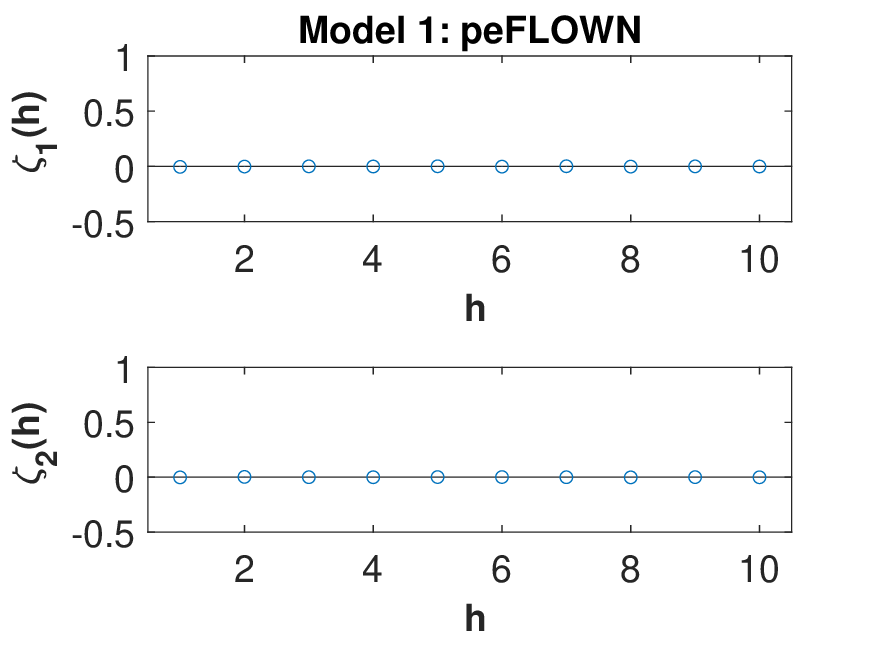}
    \includegraphics[width=0.32\linewidth]{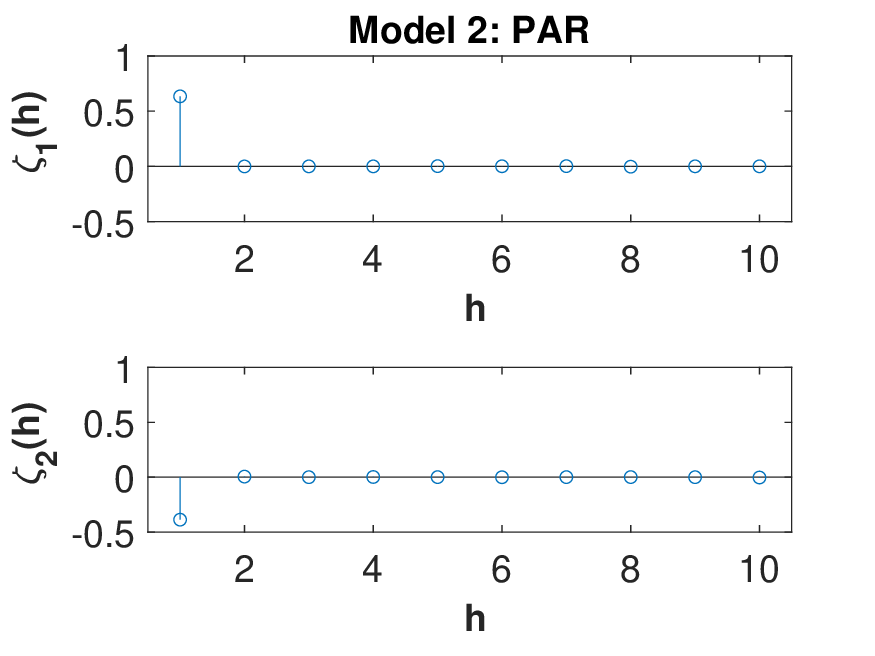}
    \includegraphics[width=0.32\linewidth]{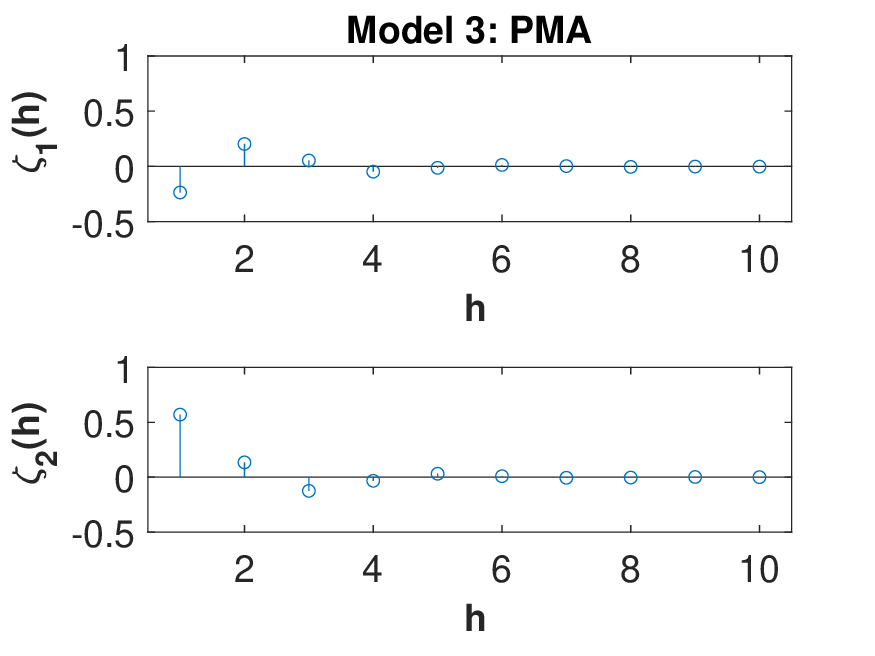}
    \caption{Values of peFLOPACF $\zeta_v(h)$ with $B=0.6$ for $v=1$ (top panels) and $v=2$ (bottom panels) for selected cyclostationary models with period $T=2$: peFLOWN (Model 1), PAR$_2(1)$ (Model 2) and PMA$_2(1)$ (Model 3). Complete specification of each model can be found in Section \ref{subsec:dists}.}
    \label{fig:sample_trajs_peflopacf}
\end{figure}

The plots of peFLOPACF $\zeta_v(h)$ (for $v=1,2$ and $h=1,2,\ldots,10$) for Models 1-3 considered in Section \ref{subsec:dists} are shown in Fig. \ref{fig:sample_trajs_peflopacf}. These results were obtained using the sample peFLOPACF $\hat{\zeta}_v(h)$ averaged over 1000 trajectories of length $NT=1000$ of given time series. We assume that $B=0.6$, so that this measure can be considered for time series with $\alpha=1.7$ (as assumed for Models 1-3). For peFLOWN sequence (Model 1), one can see that all values of peFLOPACF are equal to 0. For PMA time series (Model 3), some non-zero values of $\zeta_v(h)$ and its decay to 0 for $h\rightarrow \infty$ can be observed. As discussed above, in this case, the "cut-off" property is present for PAR model (Model 2); for $h>p$ (here, $p=1$), we have $\zeta_v(h) = 0$.

\section{Application of the peFLOACF and peFLOPACF}

In this section, we present examples of applications of the introduced peFLOACF and peFLOPACF measures in the analysis of FLOC-cyclostationary time series. First, the procedure for testing of dependence is introduced. Then, the algorithms for identifying the orders of PAR and PMA models are proposed. The presented methods are generalizations of well-known approaches designed for stationary and PC time series.

\subsection{Testing of dependence for FLOC-cyclostationary time series}
 \label{subsec:portmanteau}

In the analysis of stationary time series, portmanteau tests based on ACF (such as the Box-Pierce \cite{box1970test} or Ljung-Box tests \cite{ljung1978test}) are well-known tools for detecting autocorrelation in data. For PC time series, portmanteau tests based on the peACF were considered in \cite{mcleod1994diagnostic}. In the literature, one can also find portmanteau tests for stationary time series with infinite variance, e.g. based on covariation \cite{gallagher2006portmanteau} or codifference \cite{rosadi2009portmanteau}. In this article, we use the peFLOACF to design a portmanteau test for detecting dependence in FLOC-cyclostationary time series. 

For a zero-mean sample $x_1,\ldots,x_{NT}$ which is assumed to correspond to a FLOC-cyclostationary time series $\{X_t\}$ (with known period $T$), the following null and alternative hypotheses are considered
\begin{equation}
    \mathcal{H}_0:~~ \{X_t\}\text{ is a peFLOWN time series}, \nonumber
\end{equation}
\begin{equation}
\mathcal{H}_1:~~ \{X_t\}\text{ is not a peFLOWN time series}.
    \nonumber
\end{equation}
Let us underline that the FLOC-cyclostationarity is a prior assumption for the considered time series $\{X_t\}$ (to allow one to use the peFLOACF) and not a subject of the proposed test. The focus is on checking whether or not there is autodependence (describable by the peFLOACF) in the analyzed data. The alternative hypothesis $\mathcal{H}_1$ is equivalent to the statement that $\{X_t\}$ is a FLOC-cyclostationary time series with autodependence. 

The proposed testing procedure is composed of $T$ simultaneous tests (called later subtests), one for each $v=1,\ldots,T$. In the subtest for given $v$, we check if the corresponding sample peFLOACF $\hat{\eta}_v(h)$ contains significant non-zero values. If this occurs for any $v$, then $\mathcal{H}_0$ is rejected. For given $v$, the following subtest statistic is considered
\begin{equation}\label{eq:kappa_v}
    \kappa_v = N \sum_{h \in H_\pm} (\hat{\eta}_v(h))^2, \quad \text{where } H_\pm = \{-h_{\max},\ldots,h_{\max}\} \setminus \{0\},
\end{equation}
for some assumed $h_{\max}$. This statistic has a similar form to the Box-Pierce statistic as presented in \cite{mcleod1994diagnostic}, with the peACF replaced by the peFLOACF. Taking into account negative lags $h$ is motivated by the asymmetry of $\eta_v(h)$ (and $\hat{\eta}_v(h)$), and in the performed experiments it proved to be more efficient than considering only positive $h$.

Since the proposed algorithm is a multiple testing procedure, to maintain the "global" significance level $c$, we apply the Bonferroni correction, setting for each subtest the significance level $\tilde{c} = c/T$. The critical region (the same for each subtest) is found using Monte Carlo simulations. The distribution of the subtest statistic is approximated by simulating a large number of trajectories corresponding to $\mathcal{H}_0$, calculating $\kappa_v$ for each trajectory and deriving the interval of atypically large values of $\kappa_v$ (larger than the empirical quantile of order $1-\tilde{c}$). Because the distribution of the subtest statistic is not derived analytically but using Monte Carlo simulations, there is no need to include $N$ in \eqref{eq:kappa_v}. However, it is present in this formula for consistency with the classical Box-Pierce statistic.

The following testing procedure can be applied for different cases; however, for clarity, from now on we restrict ourselves to the $\alpha$-stable case. We assume that $\{X_t\}$ is a sequence of symmetric $\alpha$-stable random variables with the same $
\alpha$. Therefore, for all peFLOACF terms in \eqref{eq:kappa_v}, we assume such $A,B$ parameters for which $A+B<\alpha$ is satisfied. As trajectories corresponding to $\mathcal{H}_0$ (for constructing the critical region), we consider i.i.d. sequences from the $\mathcal{S}(\alpha,1)$ distribution. Since the sample peFLOACF $\hat{\eta}_v(h)$ is calculated on the standardized time series values, the $\sigma$ parameter is here not relevant. Let us also mention that in the following procedure we construct only one critical region (for $v=1$) and use it in subtests for all $v$.

The proposed testing procedure for verifying $\mathcal{H}_0$ for the analyzed zero-mean sample $x_1,\ldots,x_{NT}$, the assumed value of $\alpha$, fixed $A,B$ parameters for peFLOACF in \eqref{eq:kappa_v} (so that $A+B<\alpha$), and the significance level $c$ is as follows:
\begin{enumerate}
    \item Find the critical region for each subtest (assuming $v=1$; the critical region constructed in this step is later used for all $v$):
    \begin{itemize}
        \item Generate $M$ i.i.d. sequences of length $NT$ from $\mathcal{S}(\alpha,1)$.
        \item For $i$-th trajectory, calculate the $\kappa_v$ value from \eqref{eq:kappa_v}, denoted as $\kappa_v^{(i)}$.
        \item Construct the critical region $(Q_{1-\tilde{c}},\infty)$, where $\tilde{c}=c/T$, and $Q_{1-\tilde{c}}$ is the empirical quantile of order $1-\tilde{c}$ of the vector $[\kappa_v^{(1)},\ldots,\kappa_v^{(M)}]$.
    \end{itemize}
    \item For each $v=1,\ldots,T$, perform the following steps:
    \begin{itemize}
        \item Using the analyzed dataset $x_1,\ldots,x_{NT}$, calculate the subtest statistic value $\kappa_v$ from \eqref{eq:kappa_v}, denoted as $\kappa_v^{(0)}$.
        \item If $\kappa_v^{(0)} \in (Q_{1-\tilde{c}},\infty)$, the $\mathcal{H}_0$ is rejected and the algorithm is terminated. 
    \end{itemize}
    \item If this step is achieved, there is no evidence to reject $\mathcal{H}_0$ hypothesis.
\end{enumerate}
Alternatively, one can perform all subtests in step 2 and check if any of them rejected $\mathcal{H}_0$; obviously, even a single rejection is sufficient to reject $\mathcal{H}_0$. This approach is more informative, as it allows one to localize the detected dependence (that is, for which $v$ does it occur), but the early stopping included in the above outline may reduce computational costs.

\subsection{PAR/PMA order identification}
 \label{subsec:order_identifi}

As discussed in Sections \ref{subsec:pefloacf} and \ref{subsec:peflopacf}, the "cut-off" properties of the introduced peFLOACF and peFLOPACF measures allow one to use them for identifying, respectively, the orders of PMA and PAR models. Building upon this idea (which has already been considered for other dependence measures and models, as mentioned before), in this section, we introduce novel procedures for this purpose. In the presented formulation of these methods, we restrict ourselves to the $\alpha$-stable case -- any PAR/PMA model is assumed to have symmetric $\alpha$-stable innovations $\{\xi_t\}$ with constant (known) $\alpha$. 

Let us first consider the problem of identifying the order of the PAR model. The introduced method for this task is based on the peFLOPACF $\zeta_v(h)$ which "cuts off" for all lags larger than the order $p$ of PAR$_T(p)$ model; that is, $\zeta_v(h)=0$ for $h>p$. Hence, a natural idea for identifying this order for the analyzed dataset is to inspect the sample peFLOPACF $\hat{\zeta}_v(h)$ and look for the largest argument that yields a significantly non-zero value. Analyzing the $\hat{\zeta}_v(h)$ function in this way for each $v=1,\ldots,T$ gives the estimated seasonal order $p(v)$. The "global" order $p$ can then be derived as the maximum of all seasonal orders $p(v)$.

Whether the value of $\hat{\zeta}_v(h)$ is close to zero (or significantly non-zero) is determined by falling within (or not) the corresponding confidence interval. It is defined as a "typical" (according to the assumed confidence level $d$) range of $\hat{\zeta}_v(h)$ values when calculated for $\{X_t\}$ being a peFLOWN time series. It is analogue to what is considered in classical order identification procedures; e.g., for AR models, the distribution (confidence intervals) of the sample PACF for white noise is analyzed. In the presented method, the confidence intervals are obtained using Monte Carlo simulations, where i.i.d. sequences of $\mathcal{S}(\alpha,1)$ distribution are used as peFLOWN model (the value of $\sigma$ parameter is not relevant). We construct a separate confidence interval for each considered $h$, but only for $v=1$; these intervals are then used for all the other $v$.

For a sample $x_1,\ldots,x_{NT}$ assumed to correspond to a PAR$_T(p)$ model $\{X_t\}$, for assumed $\alpha$, fixed $B$ (such that $B<\alpha-1$; recall that $A=1$ in the peFLOPACF) and the confidence level $d$, the proposed procedure of PAR model order identification is composed of the following steps:
\begin{enumerate}
    \item Find the confidence intervals for $\hat{\zeta}_v(h)$ for each $h\in H_+ = \{1,\ldots,h_{\max}\}$ (assuming $v=1$; the confidence intervals constructed in this step are later used for all $v$):
    \begin{itemize}
        \item Generate $M$ i.i.d. sequences of length $NT$ from $\mathcal{S}(\alpha,1)$.
        \item For $i$-th trajectory, calculate $\hat{\zeta}_v(h)$ (denoted as $\hat{\zeta}_v^{(i)}(h)$) for each $h\in H_+$.
        \item For each $h\in H_+$, construct the confidence interval $(Q_{(1-d)/2}(h),Q_{1-(1-d)/2}(h))$, where $Q_{a}(h)$ is the empirical quantile of level $a$ of the vector $[\hat{\zeta}_v^{(1)}(h),\ldots,\hat{\zeta}_v^{(M)}(h)]$.
    \end{itemize}
    \item For each $v=1,\ldots,T$, perform the following steps:
    \begin{itemize}
        \item Using the analyzed dataset $x_1,\ldots,x_{NT}$, for each $h\in H_+$, calculate $\hat{\zeta}_v(h)$ denoted as $\hat{\zeta}_v^{(0)}(h)$.
        \item Find seasonal order $p(v)$ as the largest lag $h$ for which $\hat{\zeta}_v^{(0)}(h)$ is outside the corresponding confidence interval:
        \begin{equation}\label{eq:condition_pv}
            p(v) = \max\{h\: : \: h \in H_+, \,\,\, \hat{\zeta}_v^{(0)}(h) \notin (Q_{(1-d)/2}(h),Q_{1-(1-d)/2}(h)) \},
        \end{equation}
        or set $p(v)=0$, if the set in \eqref{eq:condition_pv} is empty (i.e. for each $h\in H_+$, $\hat{\zeta}_v^{(0)}(h)$ is inside the confidence interval).
    \end{itemize}
    \item As the "global" order $p$, set $p = \max_{v=1,\ldots,T} p(v)$. 
\end{enumerate}

The proposed procedure for identifying the order of PMA$_T(q)$ model has a very similar form. Here, the underlying dependence measure is the peFLOACF, which, as mentioned, has a "cut-off" property for PMA time series, that is, $\eta_v(h)=0$ for $|h|>q$. Recall that, in this case, negative lags are also considered. Similarly as before, the introduced method is composed of $T$ steps where in each the seasonal order $q(v)$ is found based on the inspection of the sample peFLOACF $\hat{\eta}_v(h)$, for $v=1,\ldots,T$. This time, the identified order is the smallest non-negative integer $k$ for which $\hat{\eta}_v(h)$ is close to zero for all $h<-k$ and $h>k$. The equivalent statement is also given below in \eqref{eq:condition_qv}. Again, for $\hat{\eta}_v(h)$ value, being close to zero is determined by falling within the corresponding confidence interval. As the "global" order $q$, we set the maximum of all seasonal orders $q(v)$.

Other technical details of the procedure for PMA models remain the same as for PAR time series, so that it can be listed in a similar manner. For a sample $x_1,\ldots,x_{NT}$ assumed to correspond to a PMA$_T(q)$ model $\{X_t\}$, for assumed $\alpha$, fixed $A,B$ (such that $A+B<\alpha$), and the confidence level $d$, the proposed procedure of order identification is composed of the following steps:
\begin{enumerate}
    \item Find the confidence intervals for $\hat{\eta}_v(h)$ for each $h\in H_\pm = \{-h_{\max},\ldots,h_{\max}\} \setminus \{0\}$ (assuming $v=1$; the confidence intervals constructed in this step are later used for all $v$):
    \begin{itemize}
        \item Generate $M$ i.i.d. sequences of length $NT$ from $\mathcal{S}(\alpha,1)$.
        \item For $i$-th trajectory, calculate $\hat{\eta}_v(h)$ (denoted as $\hat{\eta}_v^{(i)}(h)$) for each $h\in H_\pm$.
        \item For each $h\in H_\pm$, construct the confidence interval $(Q_{(1-d)/2}(h),Q_{1-(1-d)/2}(h))$, where $Q_{a}(h)$ is the empirical quantile of level $a$ of the vector $[\hat{\eta}_v^{(1)}(h),\ldots,\hat{\eta}_v^{(M)}(h)]$.
    \end{itemize}
    \item For each $v=1,\ldots,T$, perform the following steps:
    \begin{itemize}
        \item Using the analyzed dataset $x_1,\ldots,x_{NT}$, for each $h\in H_\pm$, calculate $\hat{\eta}_v(h)$ denoted as $\hat{\eta}_v^{(0)}(h)$.
        \item Find seasonal order $q(v)$:
        \begin{equation}\label{eq:condition_qv}
            q(v) = \max\{k=|h|\: : \: h \in H_\pm, \,\,\, \hat{\eta}_v^{(0)}(h) \notin (Q_{(1-d)/2}(h),Q_{1-(1-d)/2}(h)) \},
        \end{equation}
        or set $q(v)=0$, if the set in \eqref{eq:condition_qv} is empty (i.e. for each $h\in H_\pm$, $\hat{\eta}_v^{(0)}(h)$ is inside the confidence interval).
    \end{itemize}
    \item As the "global" order $q$, set $q = \max_{v=1,\ldots,T} q(v)$. 
\end{enumerate}

\section{Simulation study}

In this section, the portmanteau test proposed in Section \ref{subsec:portmanteau} and procedures for identifying PAR/PMA model order presented in Section \ref{subsec:order_identifi} are validated using Monte Carlo simulations. 

\subsection{Testing of dependence for FLOC-cyclostationary time series}

The performance of the proposed portmanteau test for autodependence is assessed by calculating its empirical power for different cases of alternative hypothesis $\mathcal{H}_1$. First, we consider the case where the analyzed sample comes from the PAR$_2(1)$ model with some assumed values of $\phi_1(1)$, $\phi_1(2)$ and with innovations $\{\xi_t\}$ being an i.i.d. sequence from the $\mathcal{S}(1.7,1)$ distribution. For such a configuration, we simulate 1000 trajectories $x_1,\ldots,x_{NT}$ and run the portmanteau test (composed of $T=2$ subtests) for each, with $h_{\max}=3$, $A=B=0.8$, and the significance level $c=0.05$ (thus, each subtest has a significance level $\tilde{c}=0.025$). To construct the critical region for each subtest, $M=10000$ simulated trajectories were used. Then, we calculate the empirical power of each subtest (fraction of trajectories where given subtest correctly rejected $\mathcal{H}_0$) and the empirical power of the entire testing procedure (fraction of trajectories where any of two subtests rejected $\mathcal{H}_0$). 

\begin{figure}
    \centering
    \includegraphics[width=0.32\linewidth]{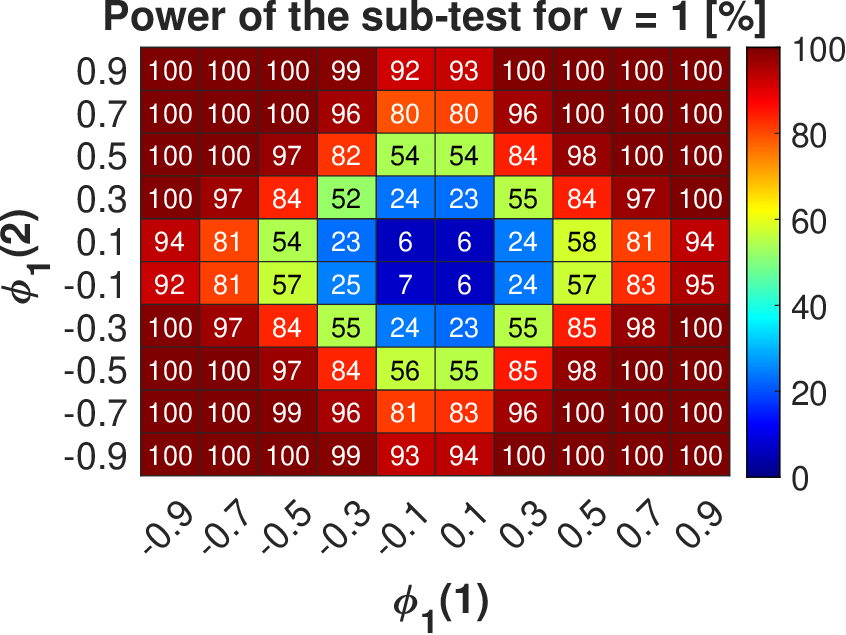}
    \includegraphics[width=0.32\linewidth]{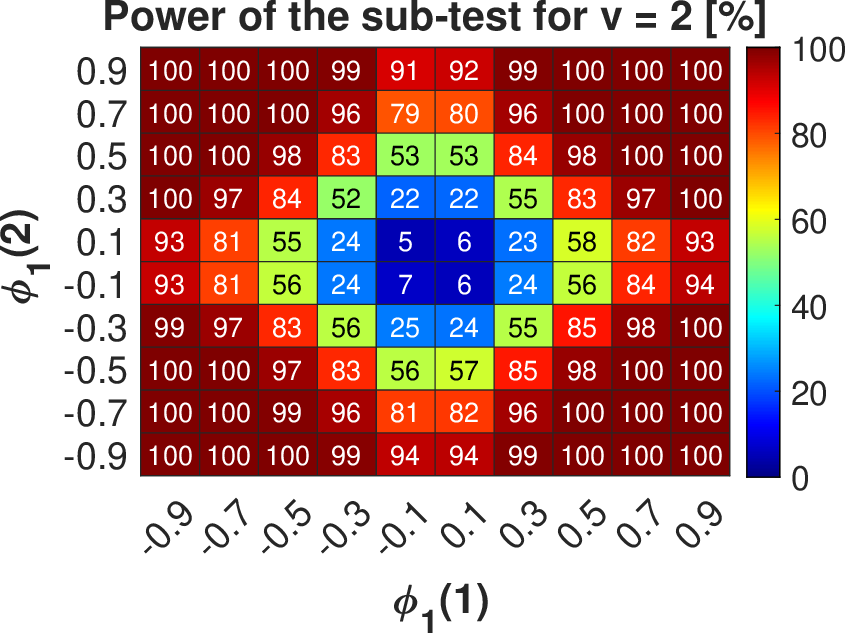}
    \includegraphics[width=0.32\linewidth]{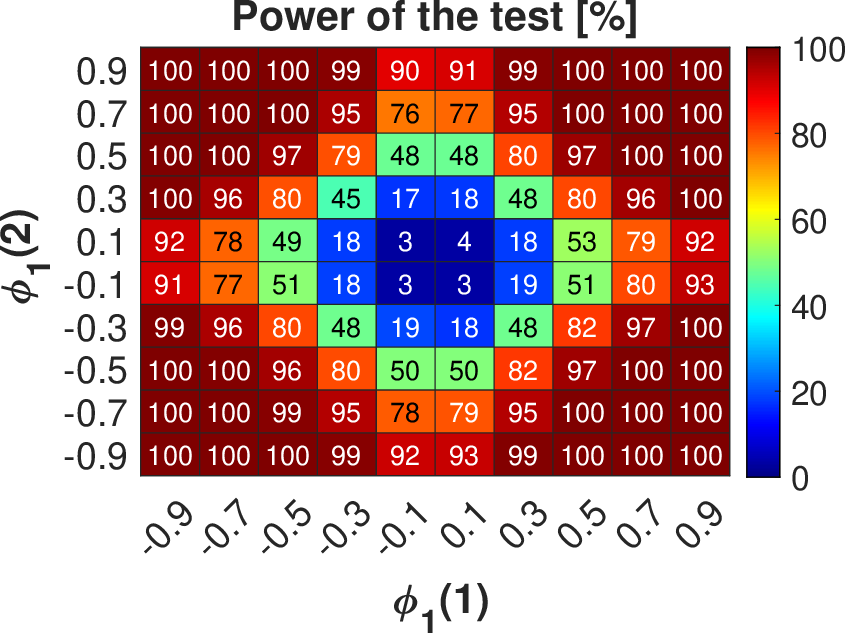}
    \caption{Empirical powers of both subtests and the entire portmanteau test for different values of $\phi_1(1)$ and $\phi_1(2)$ in PAR$_2(1)$ model in alternative hypothesis $\mathcal{H}_1$, calculated for trajectories of length $NT=100$.}
    \label{fig:[portmanteau_PAR_100]}
\end{figure}

\begin{figure}
    \centering
    \includegraphics[width=0.32\linewidth]{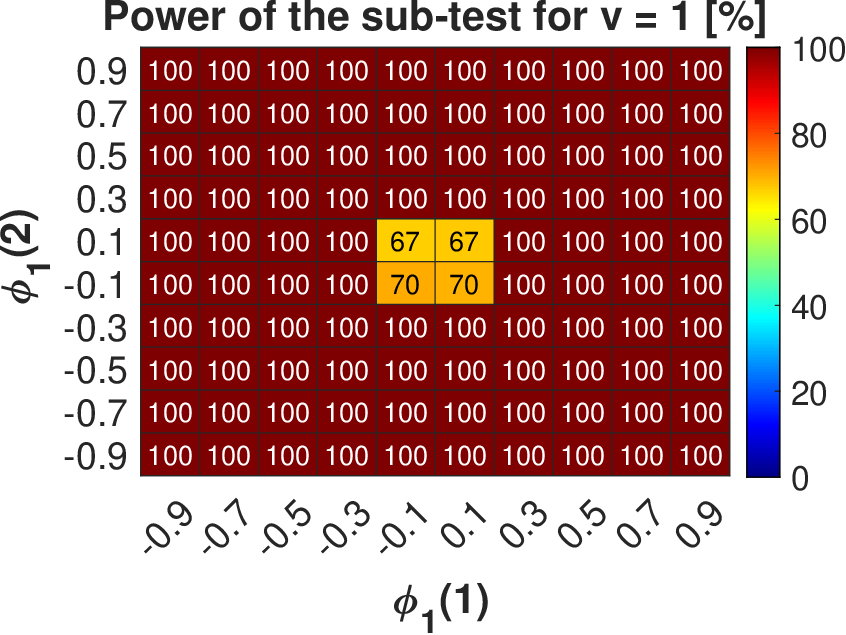}
    \includegraphics[width=0.32\linewidth]{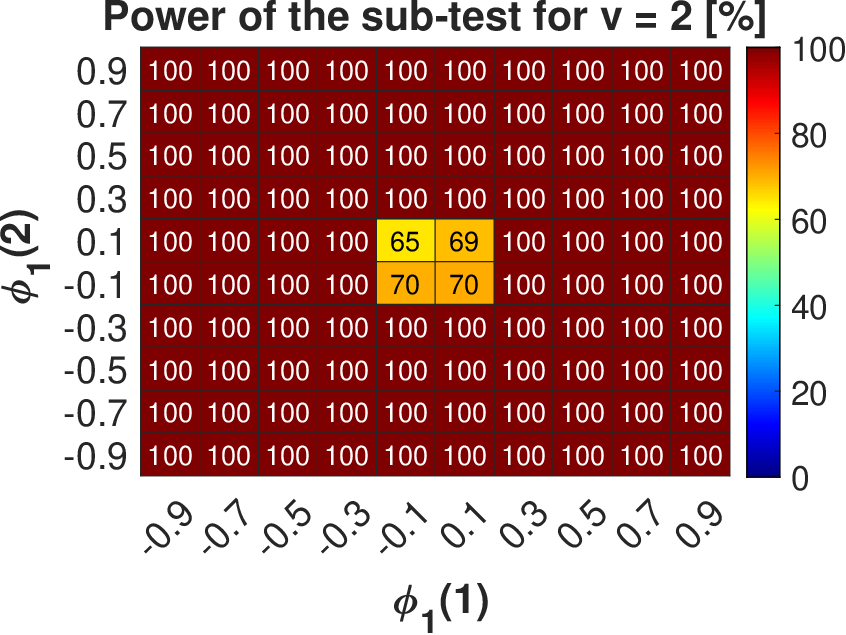}
    \includegraphics[width=0.32\linewidth]{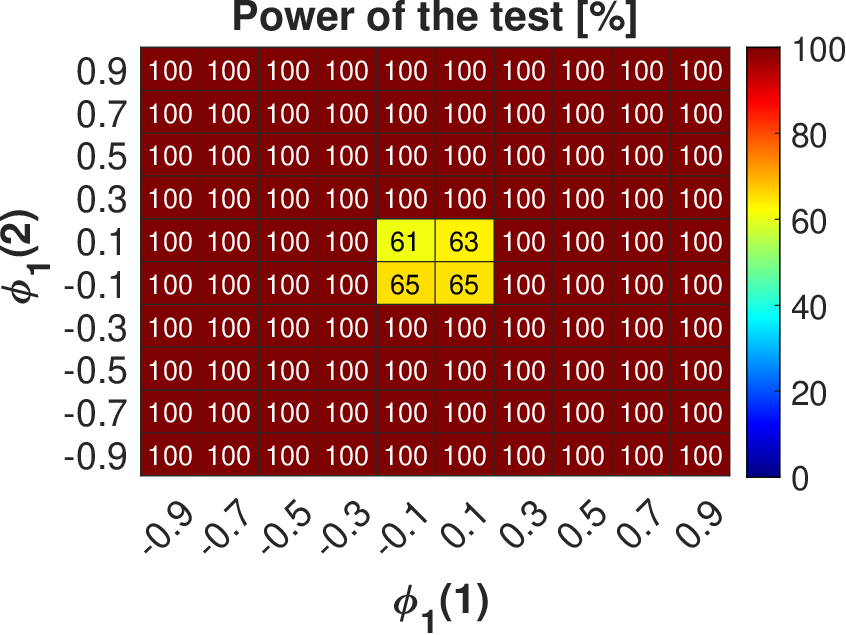}
    \caption{Empirical powers of both subtests and the entire portmanteau test for different values of $\phi_1(1)$ and $\phi_1(2)$ in PAR$_2(1)$ model in alternative hypothesis $\mathcal{H}_1$, calculated for trajectories of length $NT=1000$.}
    \label{fig:[portmanteau_PAR_1000]}
\end{figure}

This experiment is performed for all pairs $\phi_1(1),\,\phi_1(2) \in \{-0.9,-0.7,\ldots,0.7,0.9\}$. We note that the values further from 0 correspond to a higher level of observable dependence (potentially easier to detect). The results obtained for $NT=100$ are illustrated in Fig. \ref{fig:[portmanteau_PAR_100]}. As expected, the further the coefficients from 0, the better the performance of the test and the perfect efficiency is achieved relatively fast. Note that these results were obtained for rather short trajectories. The analogous results for $NT=1000$, presented in Fig. \ref{fig:[portmanteau_PAR_1000]}, show a significant improvement in efficiency --- it is perfect except for the cases of coefficients equal to -0.1 or 0.1.

\begin{figure}
    \centering
    \includegraphics[width=0.32\linewidth]{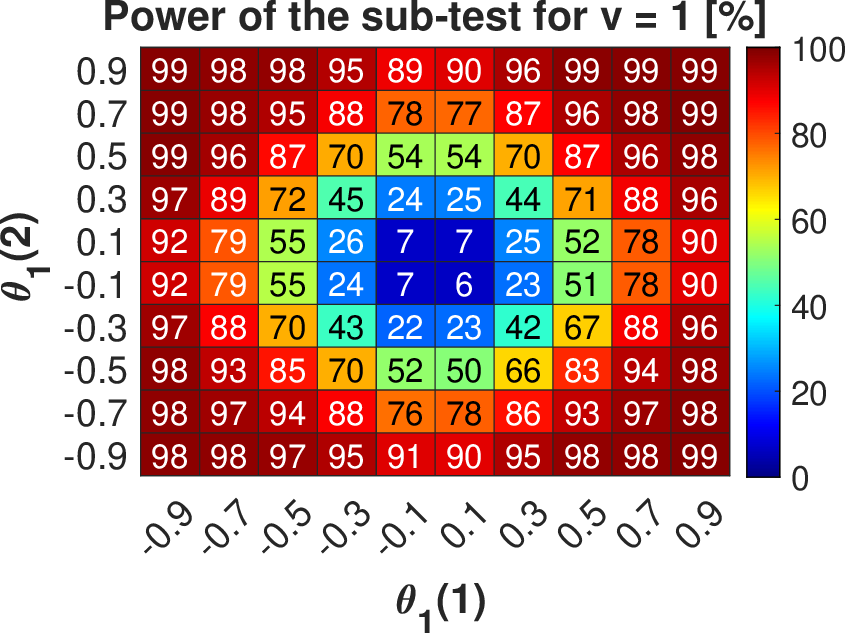}
    \includegraphics[width=0.32\linewidth]{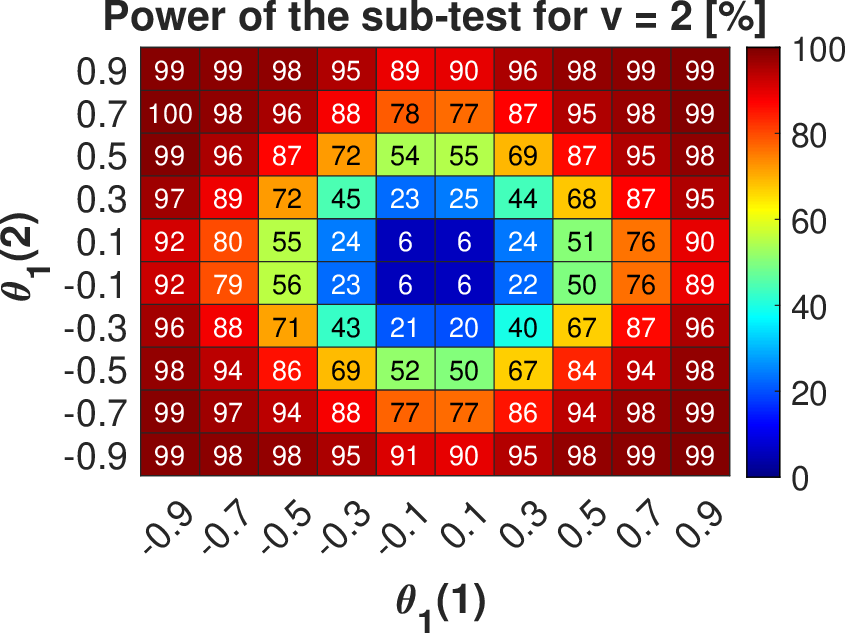}
    \includegraphics[width=0.32\linewidth]{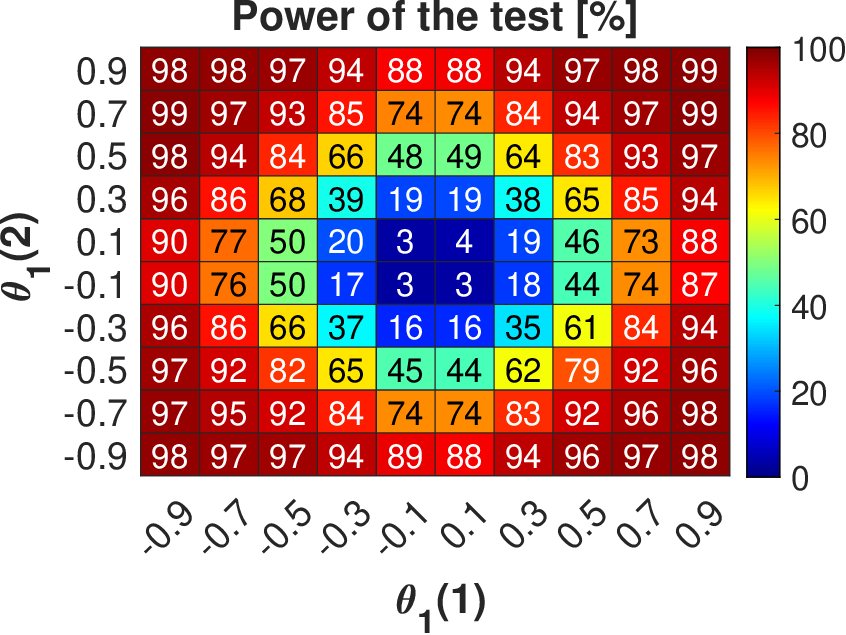}
    \caption{Empirical powers of both subtests and the entire portmanteau test for different values of $\theta_1(1)$ and $\theta_1(2)$ in PMA$_2(1)$ model in alternative hypothesis $\mathcal{H}_1$, calculated for trajectories of length $NT=100$.}
    \label{fig:[portmanteau_PMA_100]}
\end{figure}

\begin{figure}
    \centering
    \includegraphics[width=0.32\linewidth]{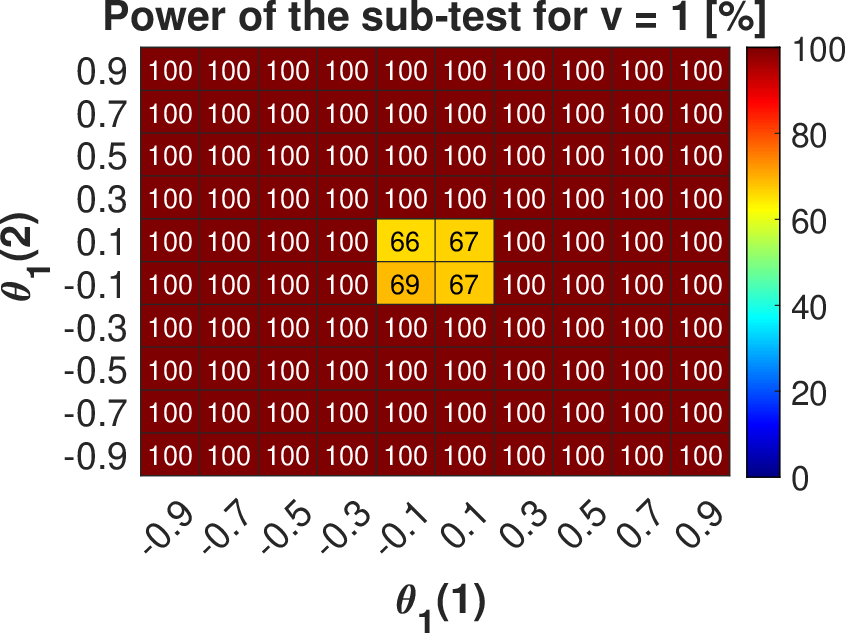}
    \includegraphics[width=0.32\linewidth]{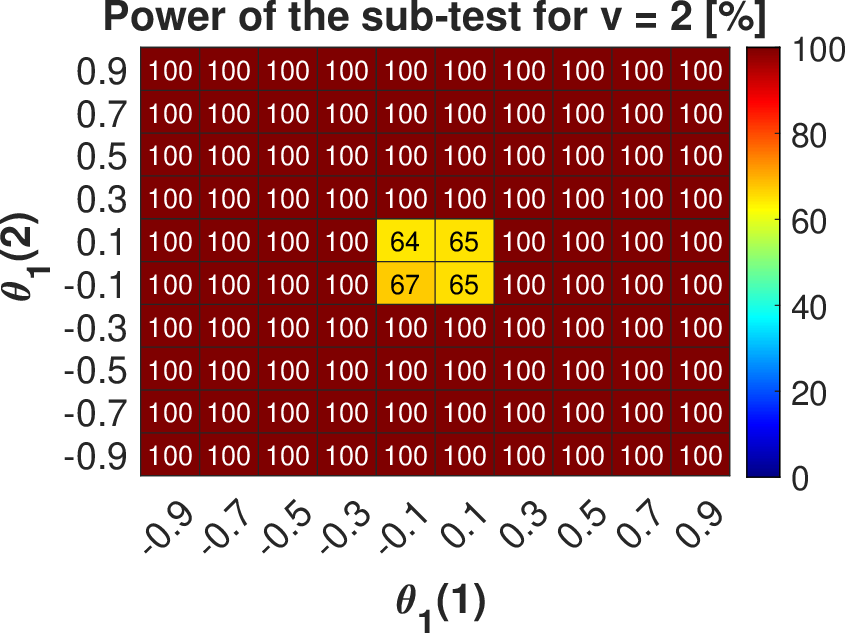}
    \includegraphics[width=0.32\linewidth]{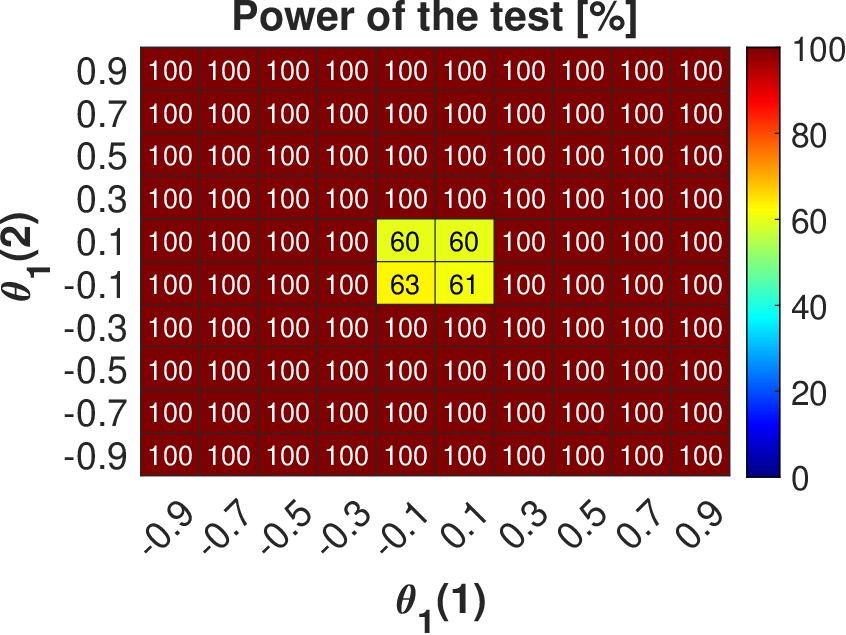}
    \caption{Empirical powers of both subtests and the entire portmanteau test for different values of $\theta_1(1)$ and $\theta_1(2)$ in PMA$_2(1)$ model in alternative hypothesis $\mathcal{H}_1$, calculated for trajectories of length $NT=1000$.}
    \label{fig:[portmanteau_PMA_1000]}
\end{figure}

To further validate the proposed portmanteau test, let us now carry out the same experiment as before with one change --- now, we assume that the alternative hypothesis $\mathcal{H}_1$ corresponds to different cases of PMA$_2(1)$ model, considering all pairs of $\theta_1(1),\,\theta_1(2) \in \{-0.9,-0.7,\ldots,0.7,0.9\}$. All other settings remain the same as in the previous analysis --- in particular, we again investigate the empirical powers of both subtests and of the entire test. The results obtained for $NT=100$ are shown in Fig. \ref{fig:[portmanteau_PMA_100]}. The general tendency is the same as for PAR models -- the more distant the coefficients, the larger the power of the test, with almost perfect efficiency achieved for the values with largest magnitude. For $NT=1000$, as illustrated in Fig. \ref{fig:[portmanteau_PMA_1000]}, the obtained powers are again equal to 100\%, except for cases with $\theta_1(1)$ and $\theta_1(2)$ equal to -0.1 or 0.1.

\subsection{PAR/PMA order identification}

Let us first analyze the order identification procedure proposed for PAR models. Similarly as in the previous section, here we consider the PAR$_2(1)$ model with assumed values $\phi_1(1)$ and $\phi_1(2)$, and with i.i.d. innovations $\{\xi_t\}$ from the $\mathcal{S}(1.7,1)$ distribution. From this model, we simulate 1000 trajectories $x_1,\ldots,x_{NT}$ and, for each, identify the PAR model order using the proposed method. Here, we use $B=0.6$, $h_{\max}=5$, confidence level $d=0.99$ and $M=10000$ simulated trajectories to construct confidence intervals. Afterwards, we calculate the percentage of cases (out of 1000 simulated trajectories of a given PAR) where the introduced algorithm correctly identified: i) $p(1)$; ii) $p(2)$; iii) both $p(1)$ and $p(2)$. Recall that, in this experiment, we have $p(1)=p(2)=1$.

\begin{figure}
    \centering
    \includegraphics[width=0.32\linewidth]{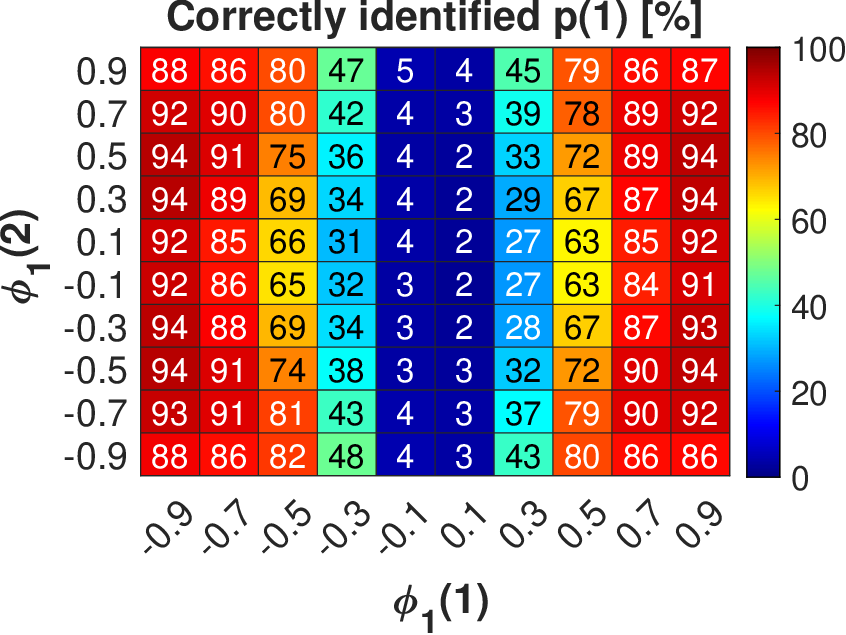}
    \includegraphics[width=0.32\linewidth]{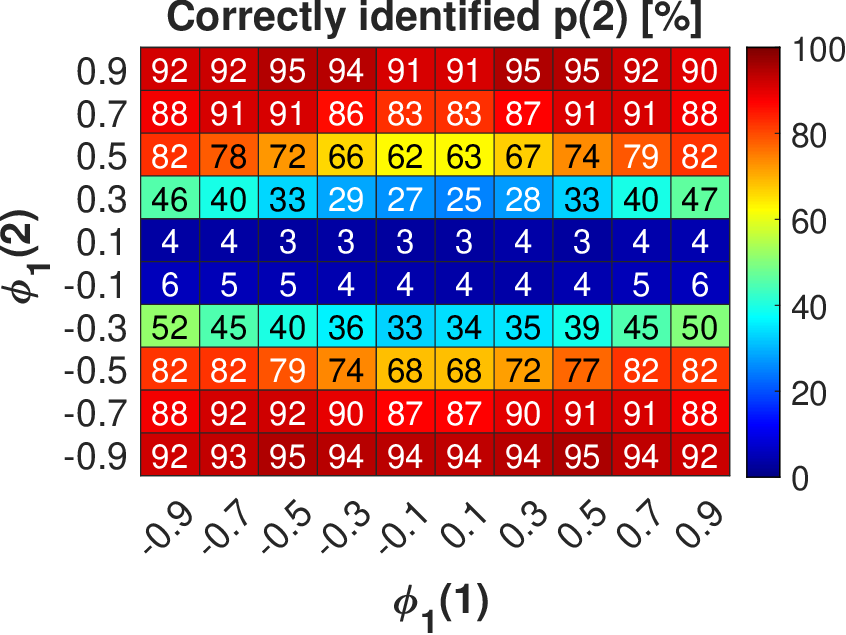}
    \includegraphics[width=0.32\linewidth]{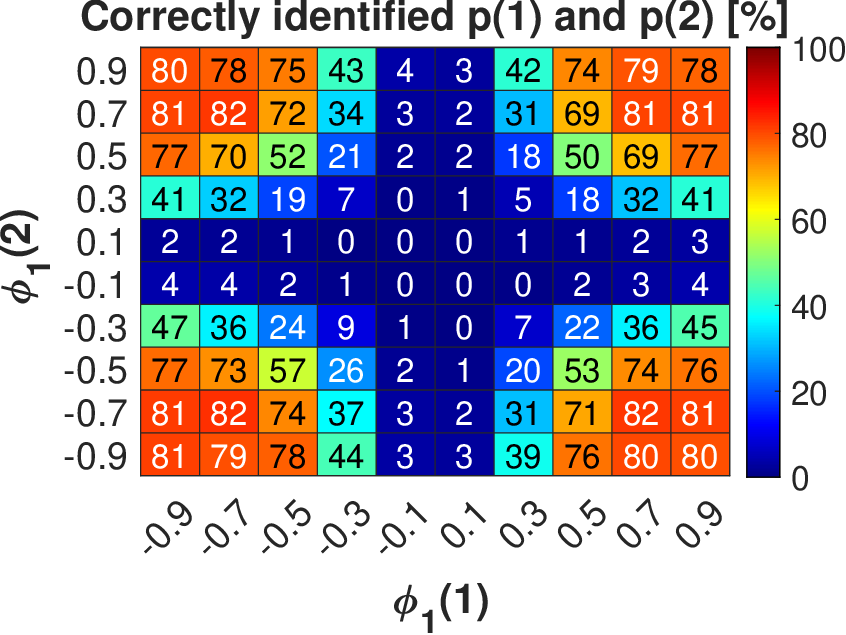}
    \caption{Percentage of cases with correctly identified $p(1)$, $p(2)$ and both $p(1),p(2)$ for different values of $\phi_1(1)$ and $\phi_1(2)$ in PAR$_2(1)$ model, for trajectories of length $NT=100$.}
    \label{fig:[order_PAR_100]}
\end{figure}

\begin{figure}
    \centering
    \includegraphics[width=0.32\linewidth]{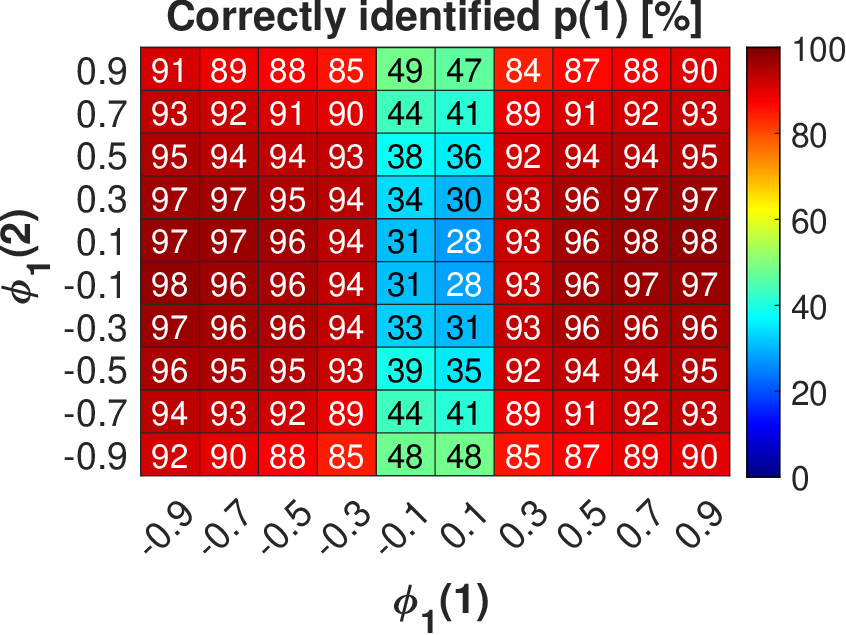}
    \includegraphics[width=0.32\linewidth]{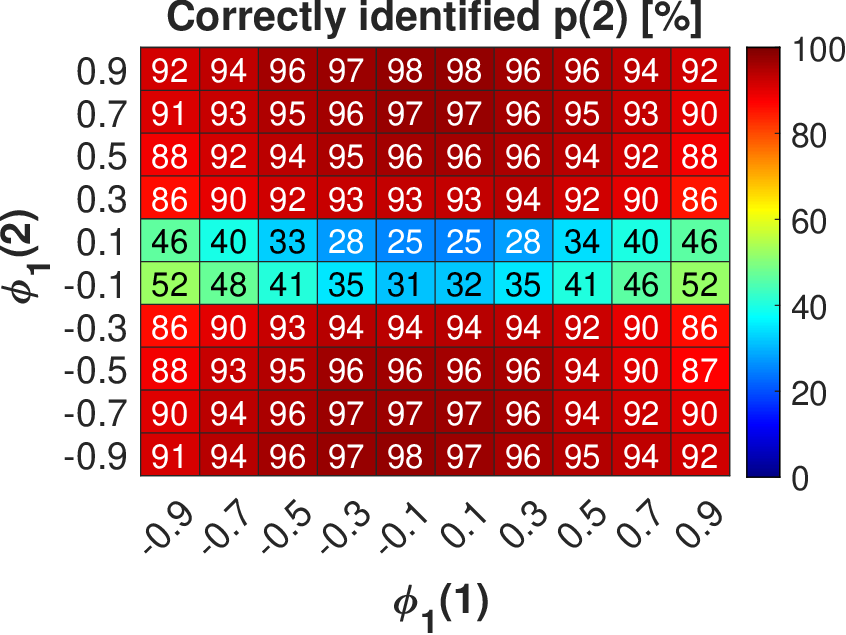}
    \includegraphics[width=0.32\linewidth]{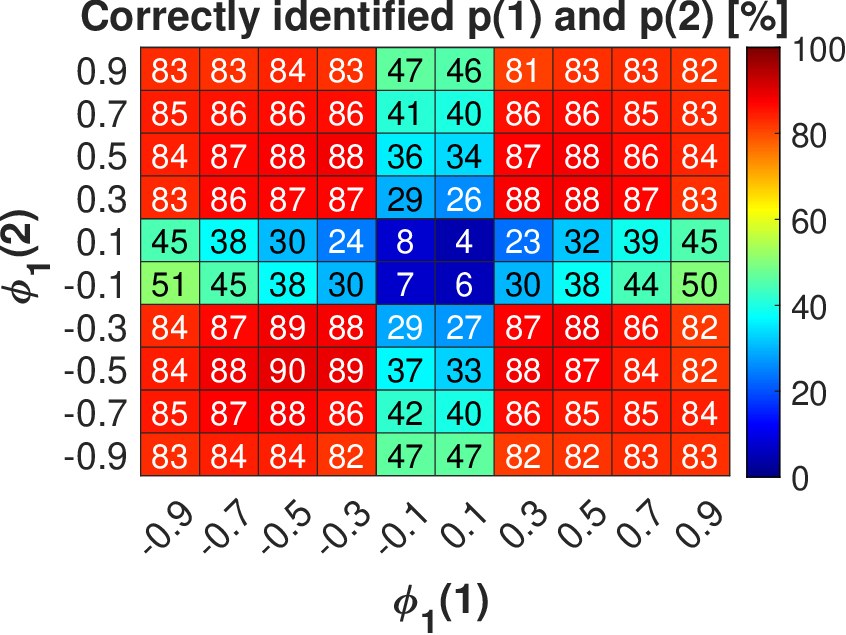}
    \caption{Percentage of cases with correctly identified $p(1)$, $p(2)$ and both $p(1),p(2)$ for different values of $\phi_1(1)$ and $\phi_1(2)$ in PAR$_2(1)$ model, for trajectories of length $NT=1000$.}
    \label{fig:[order_PAR_1000]}
\end{figure}

Again, we consider all pairs $\phi_1(1),\,\phi_1(2) \in \{-0.9,-0.7,\ldots,0.7,0.9\}$, and two sample lengths $NT=100$ and $NT=1000$. The results of this experiment for $NT=100$ are presented in Fig. \ref{fig:[order_PAR_100]}. Most importantly, one can observe that the further away from 0 the value of $\phi_1(1)$ is, the better the efficiency of the $p(1)$ identification; analogous pattern holds also for $\phi_1(2)$ and $p(2)$. Although the performance obtained for cases with coefficients of larger magnitude is acceptably high, for values close to 0 it is very poor. In such cases, most often, the proposed method returns the order equal to 0 instead of 1. However, for longer samples, this effect is greatly reduced, as illustrated in Fig. \ref{fig:[order_PAR_1000]} for $NT=1000$. Here, one can clearly see the improvement of efficiency in cases where the coefficients are close to 0. Although for coefficients equal to -0.1 or 0.1 this performance is still relatively poor, additional studies indicated that it would further improve for increased sample length.

Next, we analogously examine the efficiency of the method introduced for identifying the order of the PMA model. This experiment is almost identical to the previous one; the only differences are the considered model and the applied order identification procedure. We consider the PMA$_2(1)$ model with some assumed values of $\theta_1(1)$, $\theta_1(2)$ coefficients and i.i.d. innovations $\{\xi_t\}$ from the $\mathcal{S}(1.7,1)$ distribution. Each pair $\theta_1(1),\, \theta_1(2)\in \{-0.9,-0.7,\ldots,0.7,0.9\}$ is checked and two sample lengths $NT=100$ and $NT=1000$ are considered. For each configuration, we simulate 1000 trajectories $x_1,\ldots,x_{NT}$ of a given model and identify the order of each one using the proposed procedure, assuming $A=B=0.8$, $h_{\max}=5$, $d=0.99$, and $M=10000$. Similarly as before, we focus on the percentage of cases where the considered method correctly selected: i) $q(1)$; ii) $q(2)$; iii) both $q(1)$ and $q(2)$. Again, we have $q(1)=q(2)=1$. 

\begin{figure}
    \centering
    \includegraphics[width=0.32\linewidth]{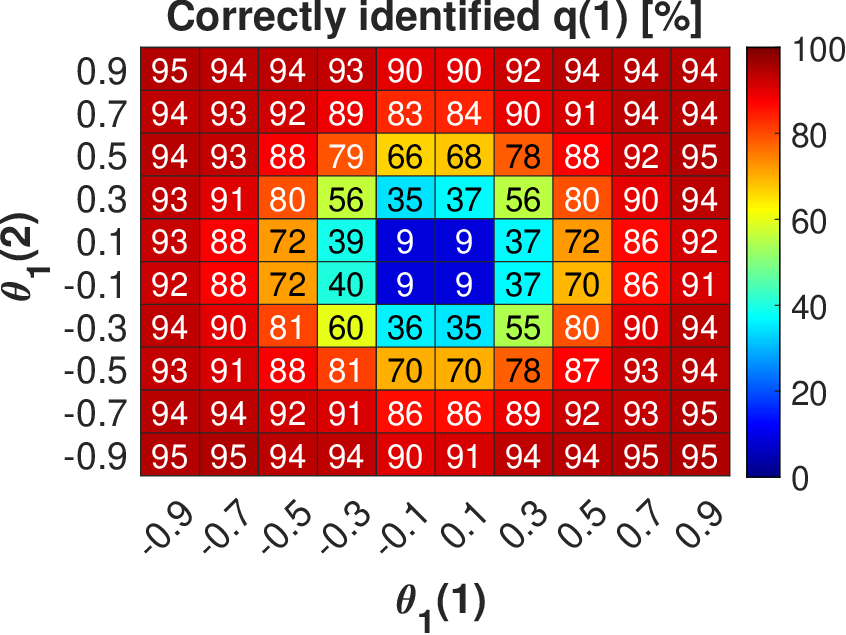}
    \includegraphics[width=0.32\linewidth]{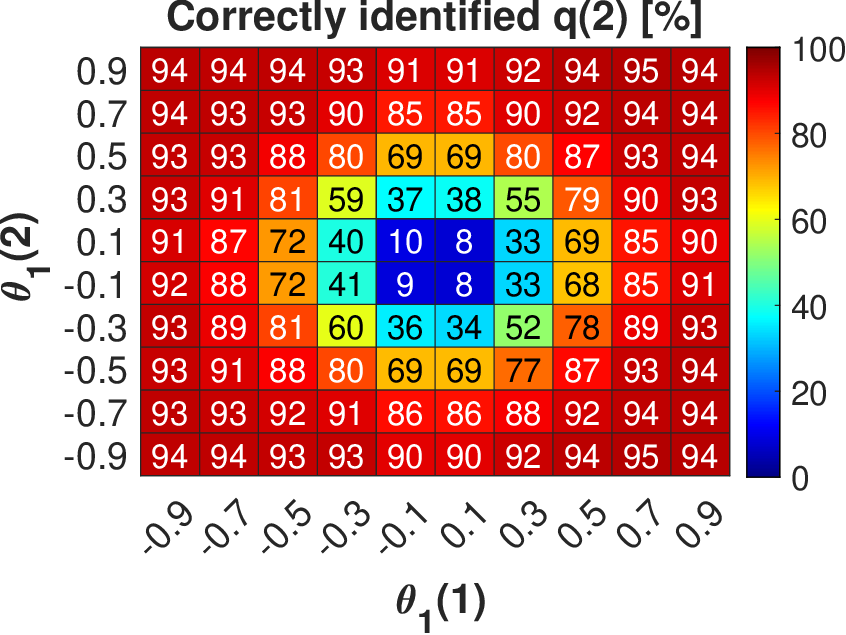}
    \includegraphics[width=0.32\linewidth]{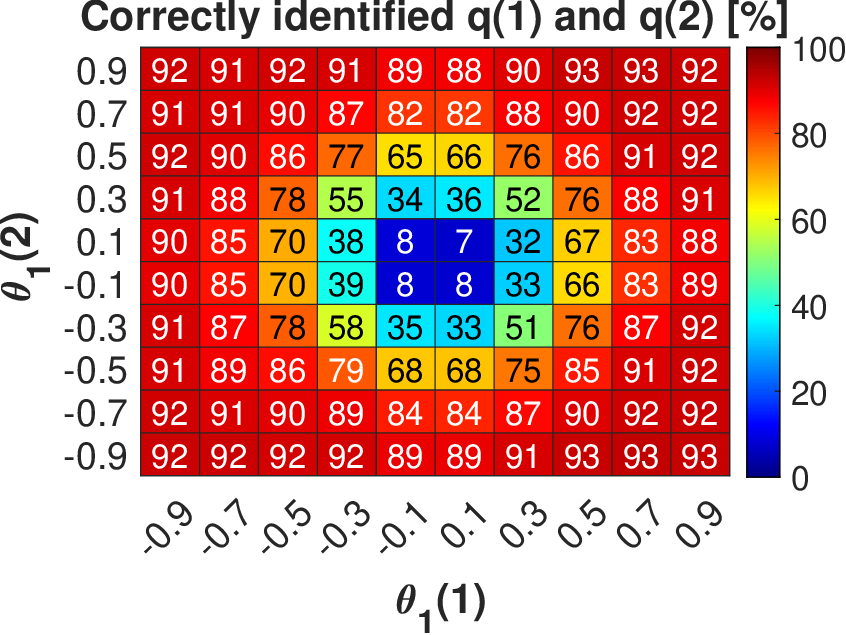}
    \caption{Percentage of cases with correctly identified $q(1)$, $q(2)$ and both $q(1),q(2)$ for different values of $\theta_1(1)$ and $\theta_1(2)$ in PMA$_2(1)$ model, for trajectories of length $NT=100$.}
    \label{fig:[order_PMA_100]}
\end{figure}

\begin{figure}
    \centering
    \includegraphics[width=0.32\linewidth]{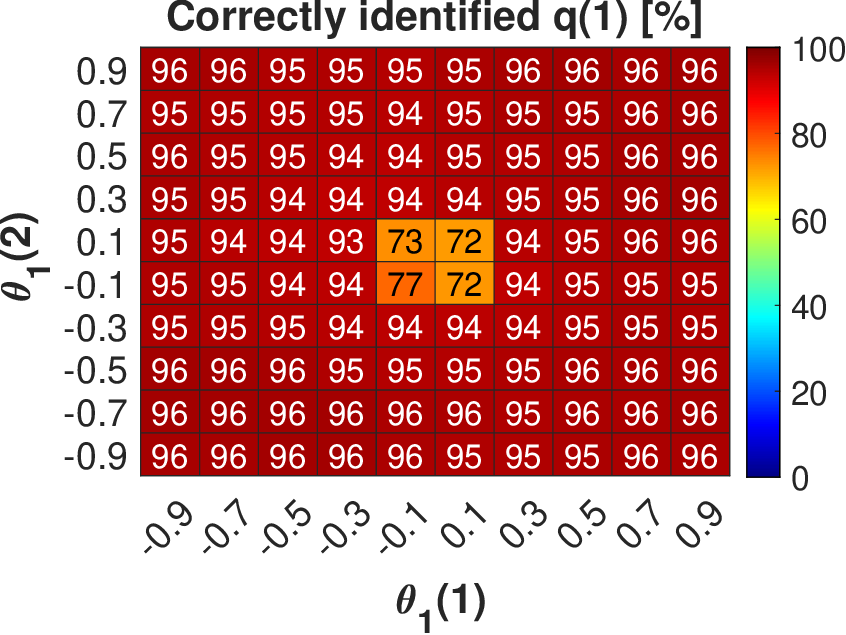}
    \includegraphics[width=0.32\linewidth]{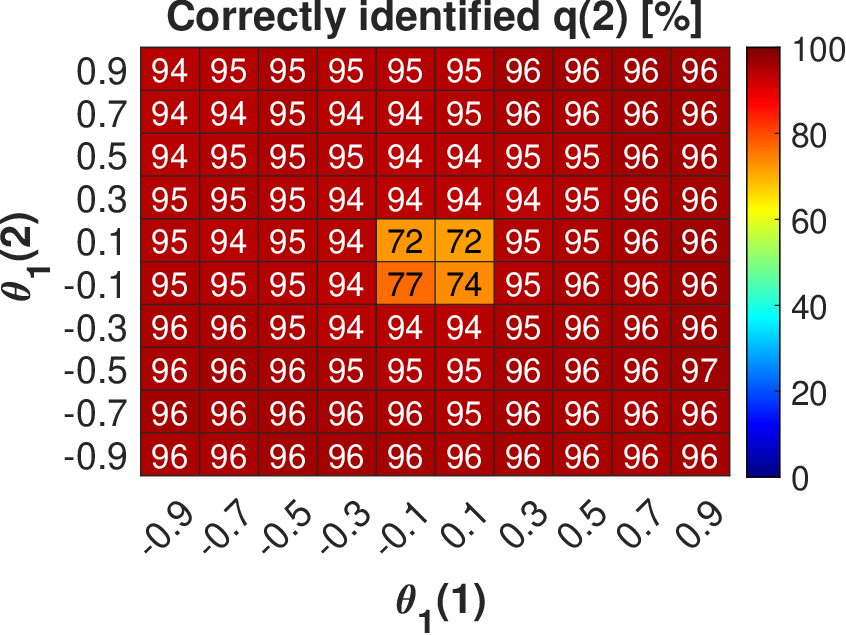}
    \includegraphics[width=0.32\linewidth]{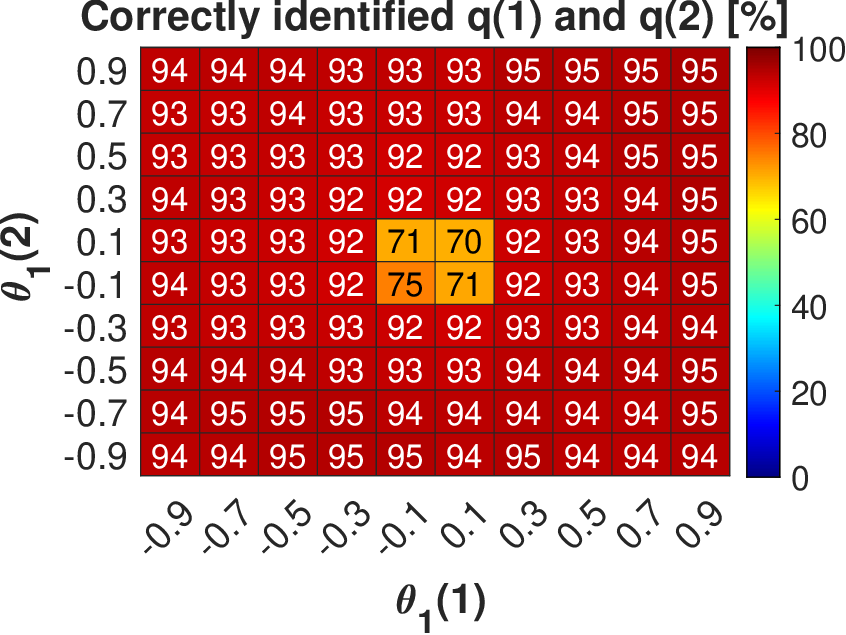}
    \caption{Percentage of cases with correctly identified $q(1)$, $q(2)$ and both $q(1),q(2)$ for different values of $\theta_1(1)$ and $\theta_1(2)$ in PMA$_2(1)$ model, for trajectories of length $NT=1000$.}
    \label{fig:[order_PMA_1000]}
\end{figure}

The results of this experiment obtained for $NT=100$ are shown in Fig. \ref{fig:[order_PMA_100]}. One can observe a slightly different behavior than in the analogous plots for PAR models. In particular, the efficiency of the $q(1)$ (or $q(2)$) identification does not seem to depend only on the $\theta_1(1)$ (or $\theta_1(2)$) value, but rather on the entire $\theta_1(1),\,\theta_1(2)$ pair. The larger the distance of both coefficients from 0, the better the performance. The lowest results were obtained for the cases where both $\theta_1(1)$ and $\theta_1(2)$ were equal to -0.1 or 0.1 (see the center of each plot), but for coefficients with larger magnitudes the efficiency is significantly better, reaching a very high level. For $NT=1000$, as illustrated in Fig. \ref{fig:[order_PMA_1000]}, a notable improvement can be observed for cases with coefficients close to 0. 

The mentioned difference in behavior between the results obtained for PAR and PMA models is also related to the fact that in the procedure proposed for PMA models negative lags are included. Additional studies indicated that including only positive lags in this method would yield a behavior similar to the one observed in the experiments for PAR models. In general, the presented results indicate the high efficiency of both proposed procedures of order identification.

\section{Real data analysis}

To illustrate the practical applicability of the proposed methodology, let us now demonstrate its usefulness in analyzing a real dataset on air pollution. The analyzed data are daily average pollutant particulate matter with a diameter smaller than 10 mm (PM$_{10}$), measured hourly in $\mu g/m^3$ at the "Vitória (center)" station located in the Great Vitória Region GVR-ES, Brazil, in the Automatic Air Quality Monitoring Network (RAMQAr). More information on the data from this source can be found in \cite{solci2020empirical} and the references therein. 

\begin{figure}
    \centering    \includegraphics[width=0.45\linewidth]{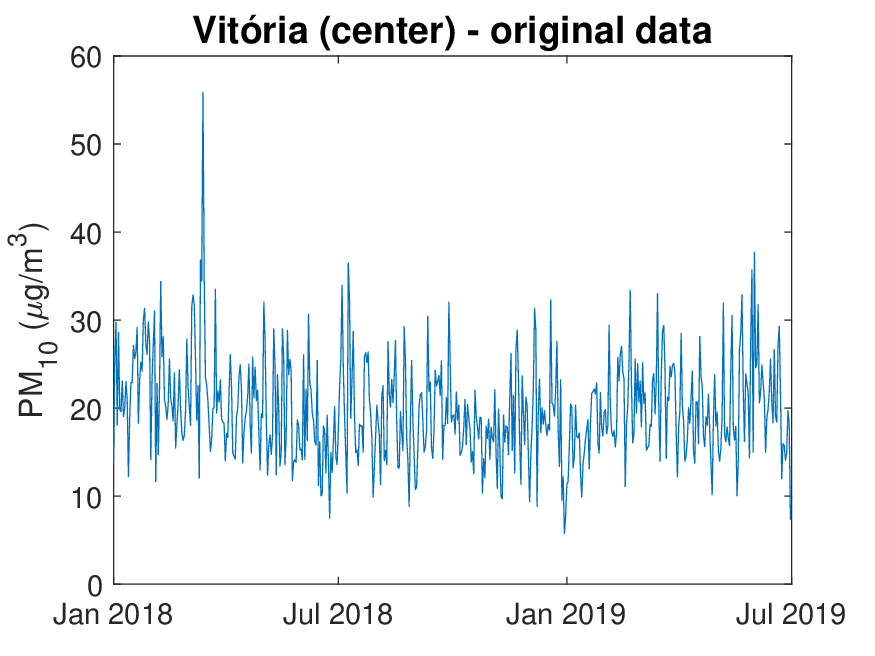}
\includegraphics[width=0.45\linewidth]{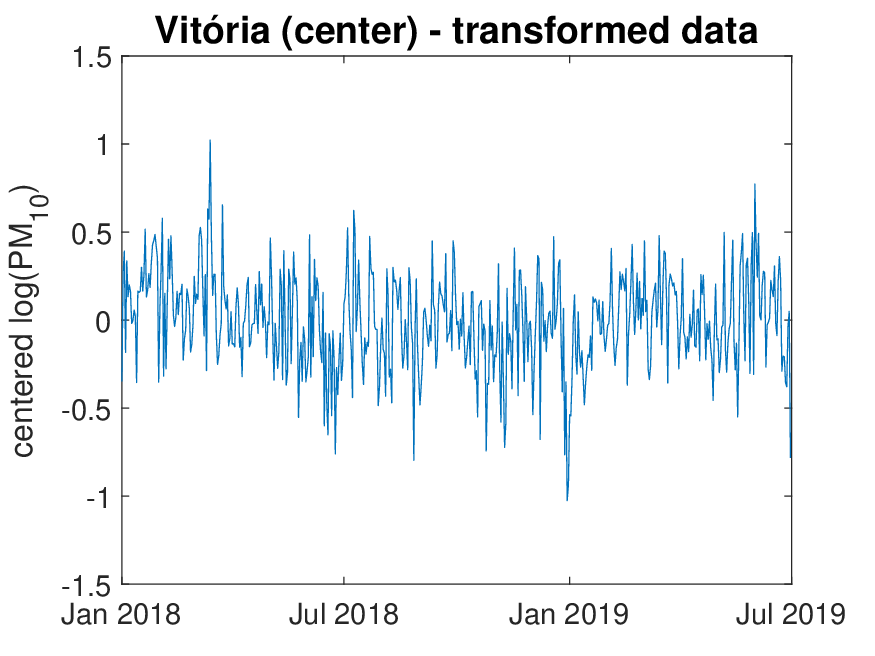}
    \caption{Analyzed dataset -- daily average PM$_{10}$ measured in the Vitória (center) station -- in the original (left) and transformed (right) form.}
    \label{fig:data_vitoria}
\end{figure}

The considered dataset was collected from 1 January 2018 to 30 June 2019 and consists of 546 observations. It is illustrated in the left panel of Fig. \ref{fig:data_vitoria}. In \cite{zulawinski2023empirical}, this dataset was analyzed using a finite-variance PAR model with additive outliers. However, for data that exhibit outliers, a different modeling approach can be considered, which assumes infinite variance of the underlying (data-generating) stochastic process. Henceforth, we follow this paradigm and assume that the analyzed dataset is a realization of a FLOC-cyclostationary and $\alpha$-stable time series. Consequently, instead of a finite-variance PAR model with additive outliers, a PAR model with symmetric $\alpha$-stable innovations is fitted. Obviously, the assumption of infinite variance requires the use of appropriate methods, such as those presented in this article. As the period, we set $T=7$ which aligns with the weekly rhythm of the considered phenomenon and was also assumed in previous analyses. 

Before the analysis, the considered dataset is preprocessed in the same way as in \cite{zulawinski2023empirical} (inspired by \cite{sarnaglia2021mregression}). The first step is the logarithmic transformation. Then, the dataset is centered by subtracting the Huber location M-estimator for each day of the week. The transformed data are illustrated in the right panel of Fig. \ref{fig:data_vitoria}. In this plot, several peaks can be observed, which motivate the application of an $\alpha$-stable distribution-based model instead of a Gaussian one. In this analysis, as the stability index, we set $\alpha=1.9$. This choice is tentatively supported by the estimate obtained using the regression-type method \cite{koutrovelis1980regression} for the entire dataset ($\alpha=1.9499$) and for each day of the week (where similar results were obtained). Being aware that the regression-type method is suitable for i.i.d. samples, we will later compare the assumed $\alpha=1.9$ with its value estimated from the residuals of the fitted PAR model.

\begin{figure}
    \centering    \includegraphics[width=\linewidth]{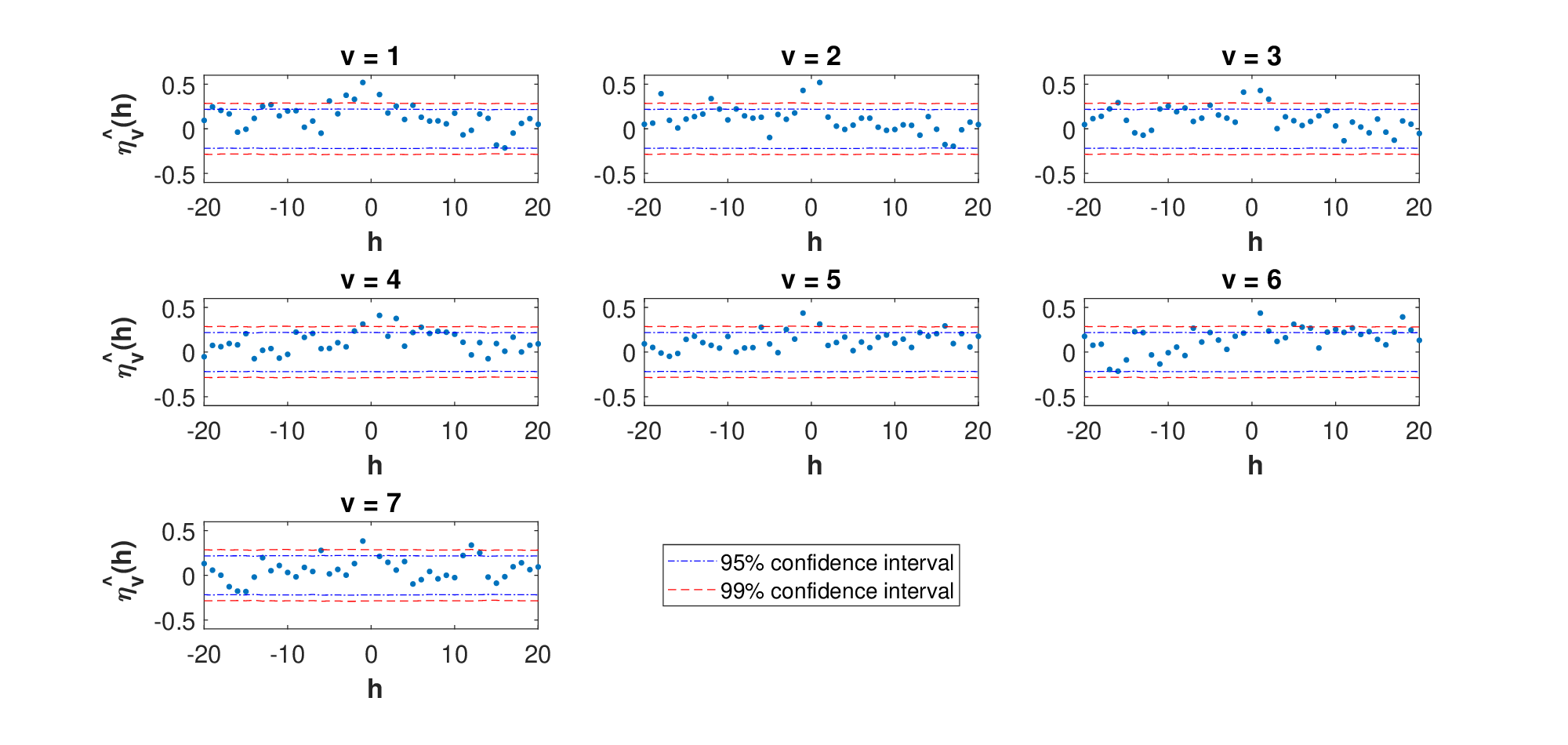}
    \caption{Sample peFLOACF $\hat{\eta}_v(h)$ (with $A=B=0.85$) for $v=1,\ldots,T$ for the analyzed dataset with confidence intervals (for i.i.d. $\mathcal{S}(1.9,1)$ sequences) at 95\% and 99\% levels.}
\label{fig:data_pefloacfs}
\end{figure}

\begin{figure}
    \centering    \includegraphics[width=\linewidth]{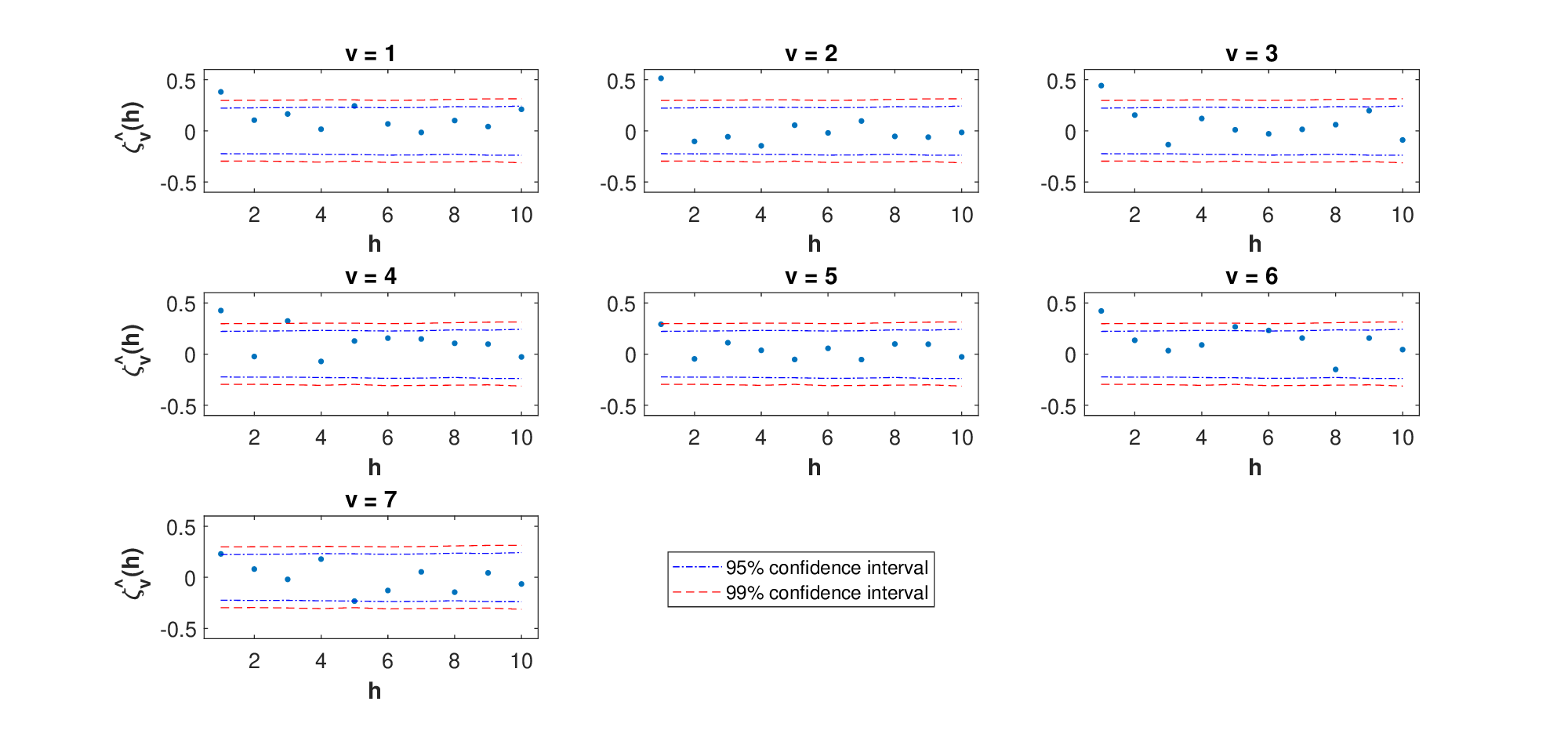}
    \caption{Sample peFLOPACF $\hat{\zeta}_v(h)$ (with $B=0.7$) for $v=1,\ldots,T$ for the analyzed dataset with confidence intervals (for i.i.d. $\mathcal{S}(1.9,1)$ sequences) at 95\% and 99\% levels.}
\label{fig:data_peflopacfs}
\end{figure}

The sample peFLOACF $\hat{\eta}_v(h)$ for $v=1,\ldots,T$ for the analyzed dataset (with $A=B=0.85$) are illustrated in Fig. \ref{fig:data_pefloacfs}. In these plots, we also include the confidence intervals for $\hat{\eta}_v(h)$ (at 95\% and 99\% levels) constructed using 100000 simulated i.i.d. sequences of length 546 from $\mathcal{S}(1.9,1)$. One can see that a short-range dependence is present in the data which supports the hypothesis that the analyzed dataset follows a PAR model. We also construct the plots of the sample peFLOPACF $\hat{\zeta}_v(h)$ (for $B=0.7$), presented in Fig. \ref{fig:data_peflopacfs}. The illustrated confidence intervals for $\hat{\zeta}_v(h)$ were built using 10000 simulated i.i.d. sequences of length 546 from $\mathcal{S}(1.9,1)$. The observed behavior is also consistent with the assumption of PAR model, including the "cut-off" pattern that will be used below to determine its order, following the procedure introduced in Section \ref{subsec:order_identifi}.  

Although the plots of the sample peFLOACF/peFLOPACF already indicated the dependence in the analyzed data, let us also confirm its presence using the portmanteau test described in Section \ref{subsec:portmanteau}, assuming $h_{\max}=20$ and the significance level $c=0.05$. The values of the subtest statistic $\kappa_v$ for $v=1,\ldots,T$, as well as the critical region, are presented in Tab. \ref{tab:portmanteau_data}. One can see that the dependence was detected for each $v$ except for $v=7$, that is, only in this case $\kappa_v$ was not inside the critical region.

\begin{table}
\centering
\begin{tabular}{| c | c | c|c|c|c|c|c|}
\hline
 $v$ & 1 & 2 & 3 & 4 & 5 & 6 & 7 \\
\hline
$\kappa_v$ & 864.3 & 648.4 & 630.0 & 607.7 & 562.1 & 905.3 & 436.4 \\
\hline\hline
\multicolumn{4}{|c|}{critical region} & \multicolumn{4}{c|}{$(458.9,\infty)$} \\ \hline
\end{tabular}
\caption{The values of the subtest statistic $\kappa_v$ and the corresponding critical region of the portmanteau test performed for the analyzed dataset.}
\label{tab:portmanteau_data}
\end{table}

As mentioned, the analyzed dataset is modeled using PAR time series. The fitting procedure begins with the identification of order using the procedure presented in Section \ref{subsec:order_identifi}. In this algorithm, we assume $d=0.99$ and $h_{\max}=10$. The obtained seasonal orders $p(v)$ are presented in Tab. \ref{tab:par_ests}. Moreover, this table includes the PAR model coefficients estimated for given $v$ using the method from \cite{zulawinski2021alternative} (denoted as $\hat{\phi}_1(v),\,\hat{\phi}_2(v),\hat{\phi}_3(v)$). The "global" order of the fitted model is $p=3$; however, it was actually identified only for $v=4$. For other days of the week, the order is equal to 1 or 0. Note that the orders identified in this step could equivalently be inferred from the visual inspection of Fig. \ref{fig:data_peflopacfs} as the largest values of $h$ (for given $v$) with the sample peFLOPACF outside the 99\% confidence interval (red dashed lines).

\begin{table}
\centering
\begin{tabular}{| c | c | c|c|c|c|c|c|}
\hline
 $v$ & 1 & 2 & 3 & 4 & 5 & 6 & 7 \\
\hline
$p(v)$ & 1 & 1 & 1 & 3 & 0 & 1 & 0 \\
\hline\hline
$\hat{\phi}_1(v)$ & 0.3621 & 0.4531 & 0.4597 & 0.3795 & 0 & 0.4842 & 0 \\
\hline
$\hat{\phi}_2(v)$  & 0 & 0 & 0 & -0.1761 & 0 & 0 & 0 \\
\hline
$\hat{\phi}_3(v)$  & 0 & 0 & 0 & 0.2851 & 0 & 0 & 0 \\
\hline
\end{tabular}
\caption{Identified orders $p(v)$ and estimated coefficients $\hat{\phi}_1(v),\,\hat{\phi}_2(v),\hat{\phi}_3(v)$ of the PAR model fitted to the analyzed dataset.}
\label{tab:par_ests}
\end{table}

\begin{figure}
    \centering    \includegraphics[width=0.45\linewidth]{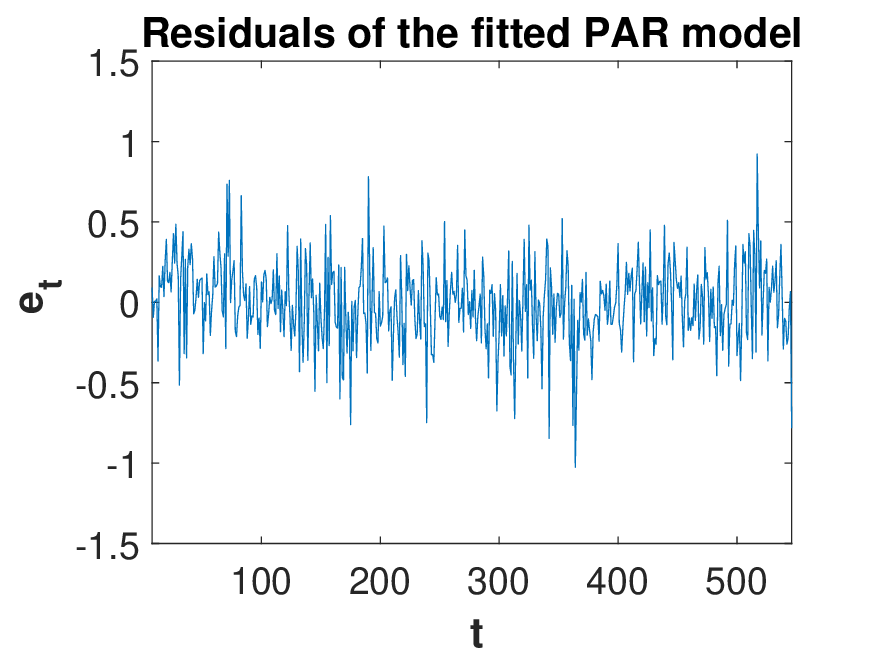}
    \caption{Time series plot of the residuals of the fitted PAR model.}
\label{fig:res_plot}
\end{figure}

To assess the goodness-of-fit of the obtained PAR model, we analyze its residuals calculated in the following way
\begin{equation}
    e_t = x_t - \hat{\phi}_1(t) x_{t-1} - \hat{\phi}_2(t) x_{t-2} - \hat{\phi}_3(t) x_{t-3},
\end{equation}
where $x_1,\ldots,x_{NT}$ is the analyzed dataset. The plot of the obtained sequence of residuals is shown in Fig. \ref{fig:res_plot}. Here, we can also observe some slightly outlying values that suggest a heavy-tailed distribution. Applying the regression-type method to the residual sequence yields the estimate $\alpha=1.89$, which is very close to the priorly assumed $\alpha=1.9$.

\begin{figure}
    \centering    \includegraphics[width=\linewidth]{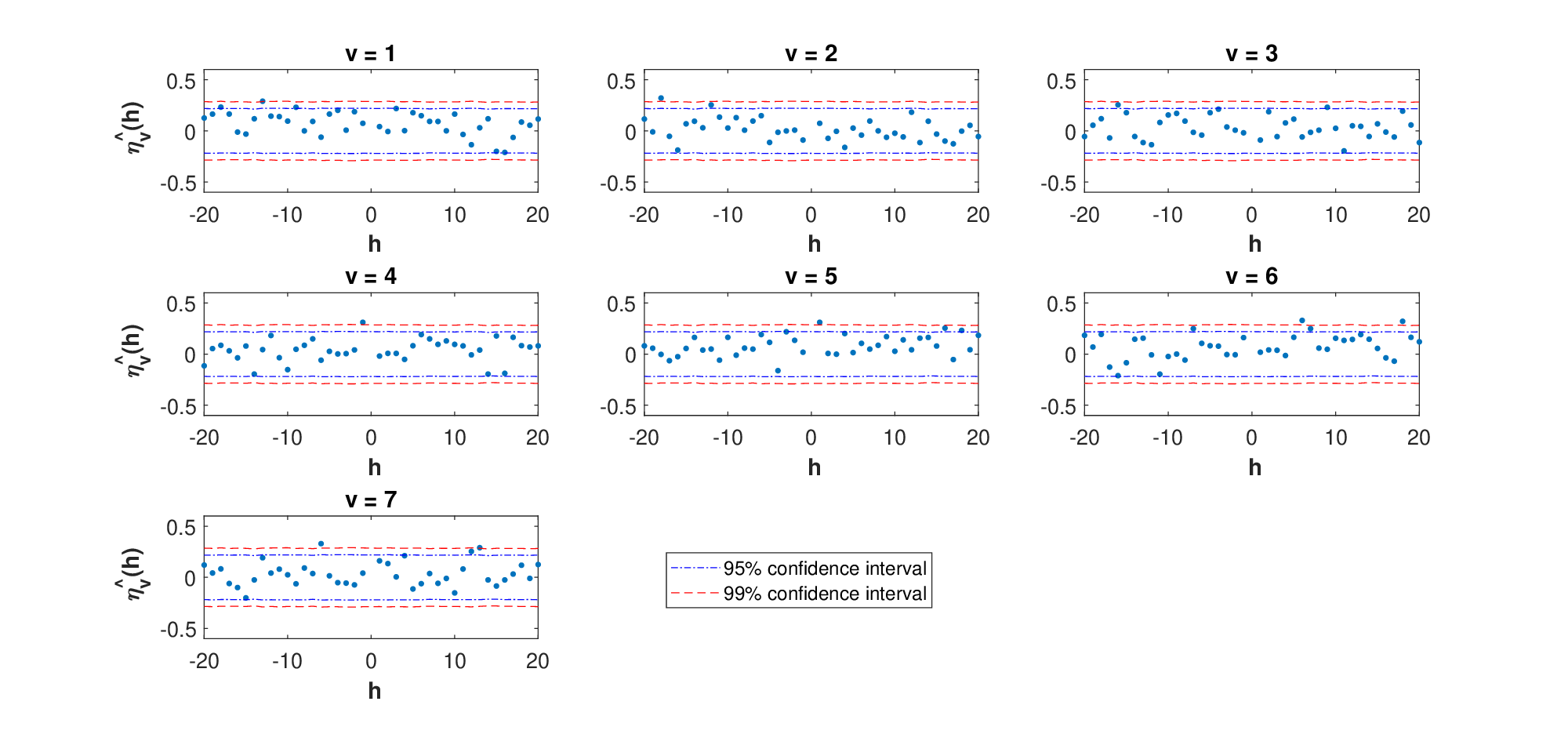}
    \caption{Sample peFLOACF $\hat{\eta}_v(h)$ (with $A=B=0.85$) for $v=1,\ldots,T$ for the residuals of the fitted PAR model with confidence intervals (for i.i.d. $\mathcal{S}(1.9,1)$ sequences) at 95\% and 99\% levels.}
\label{fig:res_pefloacfs}
\end{figure}

\begin{figure}
    \centering    \includegraphics[width=\linewidth]{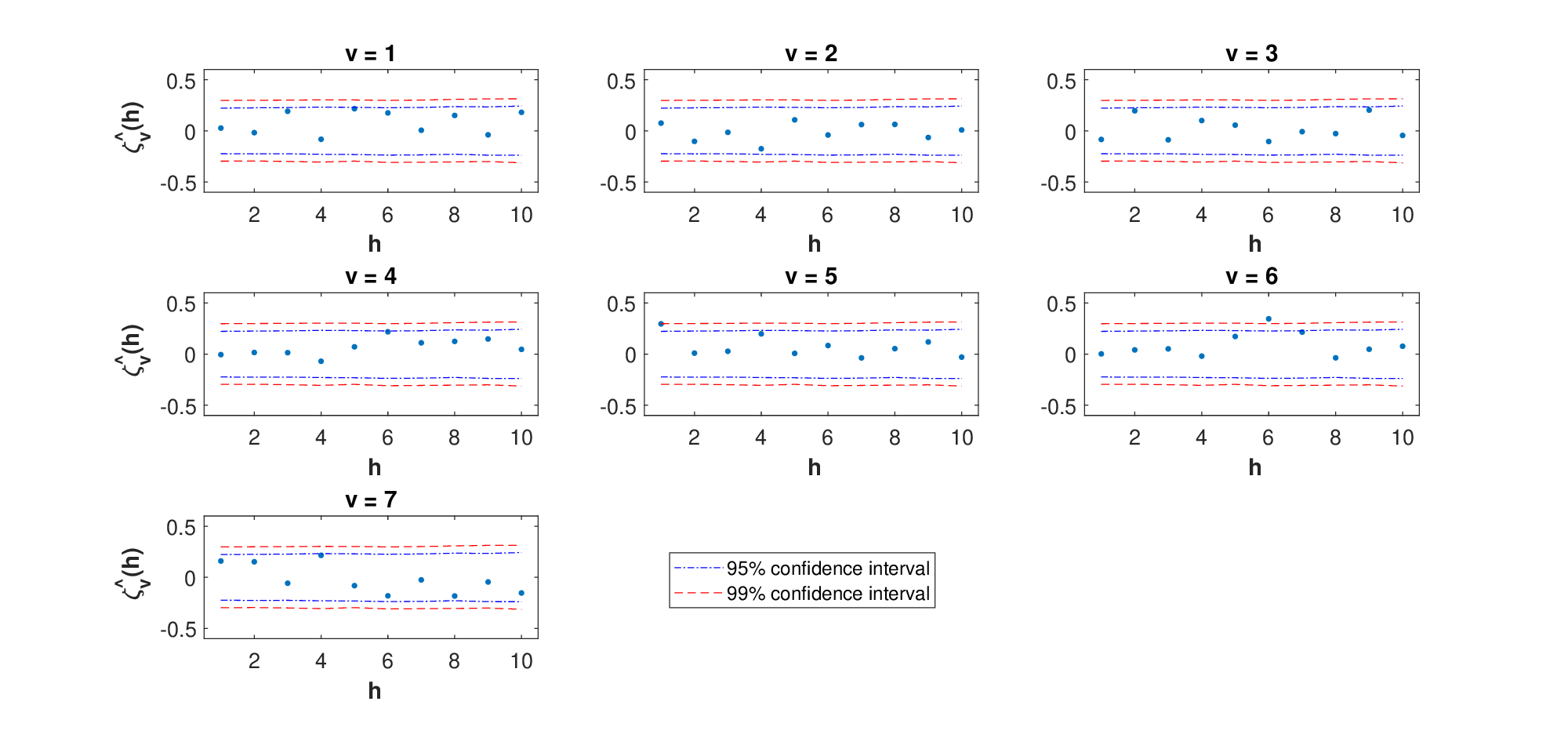}
    \caption{Sample peFLOPACF $\hat{\zeta}_v(h)$ (with $B=0.7$)  for $v=1,\ldots,T$ for the residuals of the fitted PAR model with confidence intervals (for i.i.d. $\mathcal{S}(1.9,1)$ sequences) at 95\% and 99\% levels.}
\label{fig:res_peflopacfs}
\end{figure}

In Fig. \ref{fig:res_pefloacfs}, the values of sample peFLOACF $\hat{\eta}_v(h)$ for the residuals are illustrated. Moreover, in Fig. \ref{fig:res_peflopacfs}, the sample peFLOPACF $\hat{\zeta}_v(h)$ for the residuals is shown. Both plots suggest that the residual sequence does not contain a significant autodependence, as expected for a properly fitted model. To further support this statement, the portmanteau test for the residuals is conducted in the same way as previously for the dataset. The subtest statistic values $\kappa_v$ for this case are listed in Tab. \ref{tab:portmanteau_res}. For all $v$, they are outside the derived critical region; therefore, there is no evidence to reject $\mathcal{H}_0$ and conclude that the residuals are autodependent.

\begin{table}
\centering
\begin{tabular}{| c | c | c|c|c|c|c|c|}
\hline
 $v$ & 1 & 2 & 3 & 4 & 5 & 6 & 7 \\
\hline
$\kappa_v$ & 403.3 & 254.6 & 292.7 & 287.1 & 354.9 & 450.2 & 320.6 \\
\hline\hline
\multicolumn{4}{|c|}{critical region} & \multicolumn{4}{c|}{$(458.9,\infty)$} \\ \hline
\end{tabular}
\caption{The values of the subtest statistic $\kappa_v$ and the corresponding critical region of the portmanteau test performed for the residuals of the fitted PAR model.}
\label{tab:portmanteau_res}
\end{table}

Obviously, the real data analysis presented could be further expanded; however, we underline that the main aim of this section is to present the practical applicability of the proposed methodology (peFLOACF, peFLOPACF, portmanteau test, order identification methods). In particular, we do not compare the presented modeling approach to the one applied for the analyzed dataset in \cite{zulawinski2023empirical}; the main focus is here put on the introduced measures and methods. As shown above, they may prove useful in real-world cases, extending the practical toolkit for the analysis of data exhibiting cyclic and heavy-tailed behavior.

\section{Conclusions}

In this article, cyclostationary processes with heavy-tailed distributions were analyzed. Because of the assumption of infinite variance, cyclostationarity could not be described using the classical PC property; thus, it was represented using FLOC, an alternative dependence measure. For the analysis of FLOC-cyclostationary time series, two new autodependence measures were proposed -- the peFLOACF and peFLOPACF. Based on these functions, novel procedures were introduced for testing dependence and PAR/PMA model order identification. The presented measures and derived methods generalize the concepts well established in the class of PC processes, based on the peACF and pePACF measures. As shown in the Monte Carlo simulation study, the methods introduced in this article are characterized by an acceptable efficiency and can be considered reliable in practical applications. The example of their use was also shown in the presented real data analysis. The methodology proposed in this article extends the previous research on cyclostationarity in time series with heavy-tailed (infinite-variance) distributions which is planned to be further developed. 

\section*{Funding}

\noindent The work is supported by the National Center of Science under the Sheng2 project No. UMO-2021/40/Q/ST8/ 00024 "NonGauMech - New methods of processing non-stationary signals (identification, segmentation, extraction, modeling) with non-Gaussian characteristics to monitor complex mechanical structures".

\section*{Acknowledgments}

\noindent The authors thank Prof. Valderio Reisen for sharing the real data analyzed in this article.

\bibliography{mybibliography.bib}
\end{document}